# Visual Verity in AI-Generated Imagery: Computational Metrics and Human-Centric Analysis

MEMOONA AZIZ, Department of Computer Science, Western University, Canada

UMAIR REHMAN, Department of Computer Science, Western University, Canada

SYED ALI SAFI, Department of Computer Science, Western University, Canada

AMIR ZAIB ABBASI, KFUPM Business School, King Fahd University of Petroleum & Minerals, Saudi Arabia

The rapid advancements in AI technologies have revolutionized the production of graphical content across various sectors, including entertainment, advertising, and e-commerce. These developments have spurred the need for robust evaluation methods to assess the quality and realism of AI-generated images. To address this growth, we conducted three studies. First, we introduced and validated a questionnaire namely Visual Verity, measuring photorealism, image quality, and text-image alignment. Second, we applied this questionnaire to assess images from AI models (DALL-E2, DALL-E3, GLIDE, Stable Diffusion) and camera-generated images, revealing that camera-generated images excelled in photorealism and text-image alignment, while AI models led in image quality. We also analyzed statistical properties, finding camera-generated images scored lower in hue, saturation, and brightness. Third, we evaluated Computational metrics' alignment with human judgments, identifying MS-SSIM and CLIP as the most consistent with human assessments. Additionally, we proposed the Neural Feature Similarity Score (NFSS) for assessing image quality. Our findings highlight the need for refining Computational metrics to better capture human visual perception, enhancing AI-generated content evaluation.

## 1 INTRODUCTION

The impressive scale and rapid evolution of AI-generated images (AGIs) have transformed the digital visual landscape, impacting industries such as entertainment, advertising, and e-commerce. In 2023 alone, the generation of AI-created images surpassed 15 billion, a volume that took traditional photography 150 years to accumulate. The daily production rate has reached approximately 34 million images, with platforms like Adobe Firefly contributing significantly to boosting one billion images generated by users in just three months since its launch [Everypixel Journal 2024]. The growing reliance on AI for image generation not only emphasizes the technological strides in the field but also highlights the critical need for reliable evaluation metrics that align with human perceptual standards. The image-generating segment of the generative AI market, valued at approximately 299.2 million USD in 2023, is expected to grow to 917.4 million USD by 2030, reflecting a compound annual growth rate of 17.4% [Tech Report 2024]. This rapid market expansion and the vast production of AI images underscore the importance of developing robust methods to assess the quality and photorealism of AI-generated content effectively [Caramiaux and Fdili Alaoui 2022; Talebi and Milanfar 2018b].

Given this dramatic increase in AI-generated content, there is a need to evaluate these images across key dimensions: photorealism, image quality, and text-image alignment. Photorealism ensures that images convincingly mimic real-world scenes, which is vital for applications such as virtual reality, gaming, and digital marketing. A survey revealed that 85% of users find photorealistic images more engaging, leading to increased interaction and satisfaction [Li et al. 2019; Shi et al. 2023]. Equally important is image quality, which includes attributes like sharpness, color fidelity, and the absence of artifacts. High image quality significantly influences user engagement and behavior in areas such as e-commerce and social media. Recent studies show that higher quality images see a 25% increase in user engagement and a 20% higher likelihood of purchase in online shopping scenarios [Russakovsky et al. 2014]. Text-image alignment is also important in contexts where images are paired with captions, such as educational materials, e-commerce, and social media. Effective alignment ensures that visual content complements and enhances textual information, improving comprehension and retention. A study indicates that properly aligned text and images can enhance user comprehension by up to 40% and increase user retention by 25% [Hinz et al. 2020].

In measuring these key dimensions using computational image metrics, previous research has categorized these metrics into pixel-based and model-based metrics. Pixel-based metrics, such as the Structural Similarity Index Measure (SSIM) and Peak Signal-to-Noise Ratio (PSNR) [Hore and Ziou 2010], have traditionally been used to assess the structural similarity and noise variations between AI-generated images and reference images [Talebi and Milanfar 2018a;

Wang et al. 2023]. However, the availability of a reference image is not always guaranteed. Although some datasets include similar camera-captured images that can be used as references[Lin et al. 2014b]. These metrics also require the images being compared to have the same structural composition, which can be challenging for AI-generated images with varying structures, object positions, and lighting conditions. As a result, when the outcomes of these metrics are compared to human evaluations, there is often a large discrepancy.

To overcome these limitations, model-based metrics have been developed [Jiang et al. 2022]. These metrics leverage the deep features or representations learned by pre-trained neural networks, such as Inception V3, VGG, or Vision Transformer [Rokh et al. 2023]. Since these metrics compute the distance between deep features rather than directly comparing image pixels, they allow for the assessment of images with different compositions and structures. Some prominent examples of model-based metrics include the Fréchet Inception Distance (FID) [Heusel et al. 2017], Learned Perceptual Image Patch Similarity (LPIPS) for photorealism and image quality [Zhang et al. 2018], and the Inception Score (IS) for image diversity and quality [Salimans et al. 2016].

Text-image alignment is also measured using model-based metrics such as Contrastive Language-Image Pretraining (CLIP) [Dayma et al. 2021] and Bidirectional Encoder Representations from Transformers (BERT) score [Devlin et al. 2018]. These metrics assess the alignment between textual descriptions and AI-generated images by evaluating how well-generated images match given text prompts. CLIP and BERT leverage deep features provided by pre-trained neural networks to provide alignment scores. However, human assessment remains necessary and important for accurately determining whether a generated image appropriately matches its caption, as only humans can fully appreciate the contextual and semantic accuracy of the text-image alignment [Ko et al. 2020].

Since the ultimate recipients of AI-generated images are humans, the assessment of both pixel-based and model-based metrics must be validated through human judgment [Lu and Zimmerman 2000]. However, current studies show significant gaps in comprehensively assessing AI-generated images from a human perspective. While some research has focused on specific dimensions—such as photorealism, image quality, or text-image alignment—these evaluations are often fragmented and lack a holistic approach. For instance, Le et al. [Lu et al. 2024] proposed HPBench, a study specifically targeting photorealism, while Ragot et al. [Ragot et al. 2020] investigated the alignment between human-written captions and AI-generated paintings. Zhou et al. [Zhou and Kawabata 2023] conducted a human study focusing on the beauty, liking, and valence of AI-generated art, and Treder et al. [Treder et al. 2022] emphasized the subjective evaluation of image quality through a human-centred study. Although these studies provide valuable insights, they do not integrate evaluations across the essential dimensions of photorealism, text-image alignment, and image quality comprehensively. Additionally, many of these studies lack rigorous statistical validation, which is essential for ensuring the reliability and accuracy of the assessments. Therefore, our study aims to address these gaps by providing a comprehensive, statistically validated evaluation across these critical dimensions. Without statistically robust methodologies, the conclusions drawn from human evaluations may not accurately reflect broader perceptions or withstand scrutiny in diverse applications [Treder et al. 2022].

Moreover, many current studies tend to focus on images generated by specific types of models, such as DALL-E2, DALL-E3, or Midjourney [Dayma et al. 2021; Treder et al. 2022]. This narrow focus limits the generalizability of findings across different AI models. Therefore, there is a critical need for research that not only addresses these key dimensions collectively but also incorporates rigorous statistical validation. Our study aims to fill this gap by providing a comprehensive and statistically validated assessment of photorealism, text-image alignment, and image quality across various AI models, including DALL-E2, DALL-E3, GLIDE, and Stable Diffusion.

Additionally, a significant challenge lies in scaling outputs from both metrics and human evaluations. Methods such as standardization or normalization can introduce biases, while non-linear regression-based methods are often perceived as black-box approaches, making them challenging to trust [Aziz et al. 2024]. To address this, our study proposes an expert evaluation of the scaling strategy known as the Interpolative Binning Scale (IBS) [Aziz et al. 2024] to ensure fair evaluation of human and metric outputs.

The contributions of this study are as follows:

(1) Design of three subjective questionnaires to assess key dimensions: photorealism, image quality, and text-image alignment.
(2) Statistical validation of subjective assessments using robust statistical tests and measures, ensuring the reliability and validity of the questionnaire items and excluding items based on these tests.
(3) Examination of pixel-based and model-based metrics by comparing their scores against human evaluations to determine their alignment with human perception.
(4) Design of the Neural Feature Similarity Score (NFSS), a metric designed to evaluate image quality with the goal of closely aligning with human judgment.
(5) Expert evaluation of the Interpolative Binning Scale (IBS) [Aziz et al. 2024] for conducting a fair evaluation of human and metric outputs.
(6) Investigation of the variance in the quality of images generated by different AI models through a comparative analysis of their performance.

The organization of the paper is as follows: Section 2 provides background and related work. Section 3 discusses Study 1, which involves the development and validation of the questionnaire on photorealism, image quality, and text-image alignment. Section 4 describes Study 2, which benchmarks human perception of photorealism, image quality, and text-image alignment against camera-generated and AI-generated images. Section 5 discusses Study 3, which benchmarks computational metrics for photorealism, image quality, and text-image alignment. Finally, Section 6 presents the conclusion.

## 2 BACKGROUND AND RELATED WORK

We begin by providing background and related works on AI generative image models, computational image metrics, subjective evaluation of human study and scaling strategies.

### 2.1 AI Generative Image Models

The field of AI-generated images has seen significant advancements, evolving from simple neural networks to sophisticated models leveraging transformers and diffusion techniques. This progress has been marked by the development of notable models like DALL-E [Ramesh et al. 2021], DALL-E 2 [Ramesh et al. 2022], DALL-E 3 [Betker et al. 2023], GLIDE [Nichol et al. 2021], and Stable Diffusion [Rombach et al. 2022], each contributing uniquely to the landscape of AI image generation.

The OpenAI's introduced its commercial DALL-E model in January 2021, which marked a significant milestone in AI image generation. They are utilizing transformer architectures, DALL-E demonstrated the capability to generate novel images from textual descriptions by blending various concepts and styles creatively. Despite its innovative approach, DALL-E's outputs were often limited in resolution and detail, reflecting the early stages of integrating language and image generation technologies [Ramesh et al. 2021].

GLIDE is another notable model developed by Nichol et al. [Nichol et al. 2021], which extends the principles of diffusion models by conditioning the image generation process on textual information. It is known for its ability to upscale images while maintaining high photorealism and caption accuracy, GLIDE has been considered a robust alternative to GANs and early transformer-based models. Its integration in models like DALL-E 2 highlights its importance in enhancing text-conditional image generation.

In April 2022, OpenAI released DALL-E 2. It is bringing substantial advancements over its predecessor DALL-E by incorporating a modified GLIDE model and CLIP embeddings. These enhancements allowed DALL-E 2 to generate higher resolution images, four times greater than DALL-E, with improved realism and accuracy. The model excelled in tasks like outpainting, inpainting, and creating variations from prompts, representing a substantial leap in generating photorealistic and contextually coherent images [Ramesh et al. 2022]. According to Ramesh et al., [Ramesh et al. 2022] the integration of CLIP embeddings significantly enhanced the model's ability to understand and accurately depict complex scenes.

The DALL-E 3 is the latest iteration introduced by OpenAI in September 2023. The DALL-E 3 further refined the integration of text and image generation. By leveraging more advanced techniques, DALL-E 3 aimed to improve the fidelity of generated images, especially in complex scenes with multiple objects. It addressed previous models' limitations by enhancing consistency across different parts of an image and aligning more closely with human language understanding. The improvements made DALL-E 3's text-to-image generation process more intuitive and effective, further pushing the boundaries of what these models can achieve [Betker et al. 2023].

Stable Diffusion is another advanced image generation model, which was developed by Rombach et al., [Rombach et al. 2022] at Stability AI. It has become notable for its high-resolution image generation capabilities. This model operates by iteratively refining Gaussian noise into detailed images, which makes it particularly effective for generating high-quality visuals. The open-source nature of Stable Diffusion has led to widespread adoption and application in various fields, including digital art and media. Scientific reviews have praised its flexibility and high-quality outputs, positioning it as a leading tool in the domain of AI-generated imagery [Rombach et al. 2022]. The figure 1 shows images generated by various generative models.

Assessing the quality, photorealism, and text-image alignment of images generated by these models is critical for various applications in fields such as digital marketing, virtual reality, entertainment, and education. Quality ensures that the images meet high standards of detail and clarity, which is essential for applications like e-commerce, where clear and detailed images can significantly influence purchasing decisions. For instance, higher-quality images have been shown to increase user engagement and conversion rates in online shopping platforms, as consumers and consumers are more likely to trust and purchase products that are visually appealing and accurately represented [Russakovsky et al. 2014].

### 2.2 Computational Image Metrics

In measuring image quality, photorealism, text-image alignment using computational image metrics, previous research has categorized these metrics into pixel-based and model-based metrics. Pixel-based metrics, such as the Structural Similarity Index Measure (SSIM) and Peak Signal-to-Noise Ratio (PSNR) [Hore and Ziou 2010], have traditionally been used to assess the structural similarity and noise variations between AI-generated images and reference images [Talebi and Milanfar 2018a; Wang et al. 2023]. However, the availability of a reference image is not always guaranteed. Although some datasets include similar camera-captured images that can be used as references [Lin et al. 2014b]. These metrics also require the images being compared to have the same structural composition, which can be challenging for AI-generated images with varying structures, object positions, and lighting conditions. As a result, when the outcomes of these metrics are compared to human evaluations, there is often a large discrepancy.

Pixel-based metrics directly take images as input and compute similarity scores. The Peak Signal-to-Noise Ratio (PSNR) is defined as:

$$\text{PSNR} = 10 \log_{10} \left( \frac{\text{MAX}^2}{\text{MSE}} \right) \quad (1)$$

where MAX is the maximum possible pixel value of the image and MSE is the Mean Squared Error between the reference and the generated image. PSNR is effective for measuring the overall noise level but is less sensitive to perceptual differences [Hore and Ziou 2010].

The Structural Similarity Index Measure (SSIM) improves upon PSNR by considering perceptual factors. SSIM is calculated as:

$$\text{SSIM}(x, y) = \frac{(2\mu_x \mu_y + C_1)(2\sigma_{xy} + C_2)}{(\mu_x^2 + \mu_y^2 + C_1)(\sigma_x^2 + \sigma_y^2 + C_2)} \quad (2)$$

where $\mu_x$ and $\mu_y$ are the mean intensities, $\sigma_x$ and $\sigma_y$ are the variances, and $\sigma_{xy}$ is the covariance of the images $x$ and $y$. Constants $C_1$ and $C_2$ stabilize the division with weak denominators. SSIM considers luminance, contrast, and structural similarity but still assumes the same structure in both images [Wang et al. 2023].

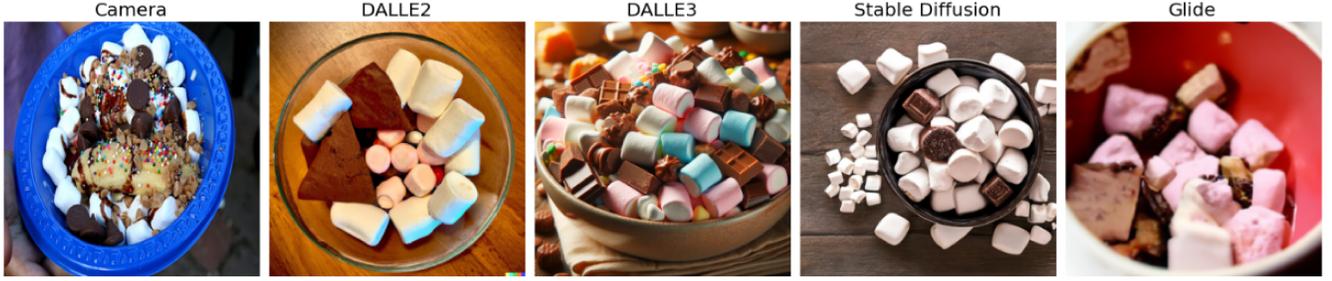

| Camera | DALLE2 | DALLE3 | Stable Diffusion | Glide |

Fig. 1. Camera captured and Images Generated by Models such as DALLE-2, DALLE-3, Stable Diffusion and Glide. The caption given to the model while generating images is "A bowl full of marshmallows, chocolate, and other delicious treats." The original image belongs to the MS-COCO dataset with ID: 000000577539.

Multi-scale structural Similarity Index Measure (MS-SSIM) extends SSIM by evaluating images at multiple scales, improving its robustness to changes in viewing conditions. MS-SSIM is computed as:

$$\text{MS-SSIM}(x, y) = \prod_{j=1}^{M} [\text{SSIM}_j(x, y)]^{\beta_j} \tag{3}$$

where $\text{SSIM}_j$ is the SSIM calculated at the $j$-th scale, and $\beta_j$ are the weights for each scale [Wang et al. 2023].

Visual Information Fidelity (VIF) measures the amount of shared information between the reference and distorted images based on natural scene statistics. VIF is defined as:

$$\text{VIF} = \frac{\sum_{j=1}^{N} I(\mathbf{C}_j; \mathbf{E}_j)}{\sum_{j=1}^{N} I(\mathbf{C}_j; \mathbf{F}_j)} \tag{4}$$

where $I(\mathbf{C}_j; \mathbf{E}_j)$ is the mutual information between the original and distorted image features, and $I(\mathbf{C}_j; \mathbf{F}_j)$ is the mutual information between the original image features and the original image itself [Sheikh et al. 2006]. VIF has been praised for its ability to align well with human visual perception, making it a robust metric for image quality assessment [Sheikh et al. 2006].

Despite their effectiveness, these traditional metrics face challenges when applied to AI-generated images due to structural variations. AI-generated images often have differences in object position, color, and lighting that pixel-based metrics are not designed to handle, leading to discrepancies between computational metric scores and human perceptual judgments. This limitation underscores the need for metrics that can better capture perceptual quality in AI-generated imagery.

To overcome these limitations, model-based metrics have been developed [Jiang et al. 2022]. These metrics leverage the deep features or representations learned by pre-trained neural networks, such as Inception V3, VGG, or Vision Transformer [Aziz et al. 2024]. Since these metrics compute the distance between deep features rather than directly comparing image pixels, they allow for the assessment of images with different compositions and structures. Some prominent examples of model-based metrics include the Fréchet Inception

Distance (FID) [Heusel et al. 2017], Learned Perceptual Image Patch Similarity (LPIPS) for photorealism and image quality [Zhang et al. 2018], and the Inception Score (IS) for image diversity and quality [Salimans et al. 2016].

The Fréchet Inception Distance (FID) is a widely used metric for assessing the photorealism and image quality of AI-generated images. It operates by comparing the statistical properties of generated images to real images using deep features extracted from a pre-trained Inception network [Aziz et al. 2024]. FID is computed by modeling the activations of a specific layer of the Inception network as multivariate Gaussian distributions for both the real and generated image sets, and then calculating the Fréchet distance between these two distributions. In most cases, the last layer before classification decision is utilized to extract deep features [Aziz et al. 2024].

The formula for FID is given by:

$$\text{FID} = \left\| \mu_r - \mu_g \right\|^2 + \text{Tr}\left( \Sigma_r + \Sigma_g - 2\sqrt{\Sigma_r \Sigma_g} \right) \tag{5}$$

where $\mu_r$ and $\Sigma_r$ are the mean and covariance of the real images, and $\mu_g$ and $\Sigma_g$ are the mean and covariance of the generated images deep features. The FID score effectively captures the difference in the distributions of real and generated images, with lower scores indicating higher similarity and thus better image quality and photorealism.

FID has been praised for its ability to correlate well with human judgments of visual quality by making it a good metric in the field of generative models. However, it is sensitive to the choice of the pre-trained network and the specific layer used for feature extraction, and it assumes that the distributions are Gaussian, which might not always hold true in practice. The Fréchet Residual Distance (FRD) is another prominent metric used to evaluate the photorealism and image quality of AI-generated images. Similar to the Fréchet Inception Distance (FID), FRD operates by comparing the statistical properties of generated images to real images. However, FRD focuses on the residuals of the images, providing a different perspective on their quality. It leverages deep features extracted from a pre-trained network, such as the ResNet, to compute these residuals [He et al. 2015].

The Kernel Inception Distance (KID) is also a model-based metric designed to address some of the limitations of FID. KID computes the squared Maximum Mean Discrepancy (MMD) between Inception representations of real and generated images, using a polynomial kernel. KID is defined as:

$$\text{KID} = \left\| \frac{1}{n} \sum_{i=1}^{n} \phi(x_i) - \frac{1}{m} \sum_{j=1}^{m} \phi(y_j) \right\|^2 \tag{6}$$

where $\phi(x)$ represents the feature map of the Inception network. KID has been shown to be more consistent than FID, especially for smaller sample sizes [Bińkowski et al. 2018].

The polynomial kernel used in KID is given by:

$$k(x, y) = (\alpha x^T y + c)^d \tag{7}$$

where $x$ and $y$ are feature vectors, $\alpha$ is a scaling factor, $c$ is a constant, and $d$ is the degree of the polynomial. The degree $d$ is a hyperparameter that influences the flexibility and complexity of the kernel. Higher degrees allow the kernel to capture more complex relationships between the features, but they also increase the risk of overfitting. Therefore, selecting an appropriate degree is important for balancing bias and variance in the metric. The use of kernels in KID allows it to handle non-linear relationships between features, providing a more nuanced comparison between real and generated images. This makes KID a robust alternative to FID, particularly when dealing with diverse and complex datasets [Aziz et al. 2024].

The Learned Perceptual Image Patch Similarity (LPIPS) metric evaluates perceptual similarity by comparing deep features extracted from multiple layers of a pre-trained network, such as VGG [Aziz et al. 2024]. LPIPS is calculated as:

$$\text{LPIPS}(x, y) = \sum_{l} \frac{1}{H_l W_l} \sum_{h,w} \|\phi_l(x)_{hw} - \phi_l(y)_{hw}\|_2^2 \tag{8}$$

where $\phi_l$ denotes the features extracted from the $l$-th layer, and $H_l$ and $W_l$ are the height and width of the feature map at layer $l$. LPIPS has been widely used for its robustness to structural variations and alignment with human perceptual judgments [Zhang et al. 2018].

The Inception Score (IS) is also a widely recognized metric for evaluating the quality and diversity of AI-generated images. This metric assesses the images by utilizing a pre-trained Inception network to classify them into different categories. The IS is computed by considering both the entropy of the conditional class distribution given a generated image and the marginal class distribution. A high inception score indicates that the generated images are both diverse and recognizable as specific classes by the Inception network.

The formula for the Inception Score is given by:

$$\text{IS} = \exp\left(\mathbb{E}_{\mathbf{x}}\left[D_{KL}(p(y|\mathbf{x}) \| p(y))\right]\right) \tag{9}$$

where $p(y|\mathbf{x})$ is the conditional probability of the class label $y$ given the generated image $\mathbf{x}$, and $p(y)$ is the marginal class distribution. The Kullback-Leibler (KL) divergence measures how much the conditional distribution diverges from the marginal distribution, with higher divergence indicating better image quality and diversity.

IS has been extensively used in the evaluation of generative models, particularly GANs, due to its simplicity and effectiveness. However, it has some limitations, such as its sensitivity to the choice of the pre-trained network and its inability to handle images that do not belong to the predefined classes of the network. Despite these drawbacks, IS remains a valuable tool for comparing the performance of different generative models [Salimans et al. 2016].

By leveraging these model-based metrics, researchers can more effectively evaluate the quality of AI-generated images, taking into account both the perceptual aspects and structural variations that traditional pixel-based metrics might overlook. This comprehensive approach helps bridge the gap between computational evaluations and human perceptual assessments by ensuring more accurate and reliable measurements of image quality. However, assessment of these metrics by human is equally important for ensuring their robustness. The Table 1 showing metrics and their applications in key dimensions of image assessment.

Text-image alignment is also measured using model-based metrics such as Contrastive Language-Image Pretraining (CLIP) [Dayma et al. 2021] and Bidirectional Encoder Representations from Transformers (BERT) score [Devlin et al. 2018]. These metrics assess the alignment between textual descriptions and AI-generated images by evaluating how well-generated images match given text prompts. CLIP utilizes a contrastive learning approach to align images and text in a shared embedding space [Dayma et al. 2021]. By training on a large dataset of images paired with their corresponding captions, CLIP learns to encode images and text into vectors that are close together if they match and far apart if they do not. The alignment score is then calculated based on the cosine similarity between these vectors. CLIP's ability to understand and match text-image pairs accurately has made it a robust tool for evaluating text-image alignment in generative models. CLIP not only improves the semantic coherence between the text and the generated image but also ensures that the visual elements depicted are relevant to the provided description [Devlin et al. 2018].

The BERT score is another metric used for evaluating text-image alignment. It leverages the deep contextual embeddings provided by the BERT model to compare the generated captions with reference captions. The BERT score calculates the similarity between the predicted and reference text by considering the embeddings from multiple layers of the BERT model. This approach captures the semantic and contextual relevance better than traditional word overlap metrics. BERT score has been particularly useful in scenarios where the generated images need to closely adhere to the descriptive text, ensuring high relevance and coherence [Devlin et al. 2018].

Both CLIP and BERT score leverage deep features provided by pre-trained neural networks to provide alignment scores. However, human assessment remains necessary and important for accurately determining whether a generated image appropriately matches its caption, as only humans can fully appreciate the contextual and semantic accuracy of the text-image alignment [Devlin et al. 2018]. Human evaluations provide an irreplaceable layer of validation by ensuring that the metrics used align well with real-world perceptual judgments.

## 2.3 Subjective Evaluations of Generative Images

Human visual perception plays a important role in image comprehension, particularly in evaluating image quality and realism. Li et al. [Li et al. 2023] introduced a comprehensive questionnaire designed to assess the general quality and fidelity of AI-generated images. The study focused on gathering human feedback on various aspects of image quality, such as sharpness, color accuracy, and overall aesthetic appeal. This questionnaire aimed to standardize the subjective evaluation process, making it easier to compare human assessments with computational metrics. The findings from their study underscored the importance of human input in validating the perceived quality of generative images, highlighting discrepancies between human judgments and algorithmic assessments [Li et al. 2023].

Table 1. Categorization of Metrics in Image Assessment Domains

| Sr No | Metric | Type | Photorealism | Image Quality | Text-Image Align. |
|---|---|---|---|---|---|
| 1 | FID [Heusel et al. 2017] | Model Based | ✓ | ✓ | |
| 2 | PSNR [Hore and Ziou 2010] | Pixel based | | ✓ | |
| 3 | KID [Bińkowski et al. 2018] | Model Based | ✓ | ✓ | |
| 4 | LPIPS [Zhang et al. 2018] | Model Based | ✓ | ✓ | |
| 5 | SSIM [Wang et al. 2023] | Pixel based | ✓ | ✓ | |
| 6 | FRD [He et al. 2015] | Model Based | ✓ | ✓ | |
| 7 | IS [Salimans et al. 2016] | Model Based | | ✓ | |
| 8 | MS-SSIM [Wang et al. 2023] | Pixel Based | ✓ | ✓ | |
| 9 | VIF [Sheikh et al. 2006] | Model Based | | ✓ | |
| 10 | CLIP [Dayma et al. 2021] | Model Based | | | ✓ |
| 11 | BERT [Devlin et al. 2018] | Model Based | | | ✓ |

Lu et al. [Lu et al. 2024] advanced the field by proposing HPBench, a benchmarking framework specifically designed for the subjective evaluation of photorealism in AI-generated images. Their research provided a structured approach to quantify photorealism, involving participants in a series of controlled experiments where they rated the realism of images generated by various AI models. The HPBench framework emphasized the need for high-fidelity images in applications like virtual reality and film production, where visual realism is paramount. By establishing benchmarks for photorealism, this study contributed significantly to the development of more sophisticated and realistic generative models [Lu et al. 2024].

Ragot et al. [Ragot et al. 2020] conducted a pivotal study that explored the alignment of painting captions with their corresponding visuals. Their research aimed to understand how well AI-generated captions matched the content and context of the images they described. By involving human participants in evaluating the accuracy and relevance of these captions, the study provided valuable insights into the capabilities and limitations of current text-to-image generation models. This work highlighted the importance of semantic coherence in generative models, especially in applications like digital storytelling and educational tools where accurate descriptions are important [Ragot et al. 2020].

Qiao et al. [Qiao and Eglin 2011] study realistic behavior in computer-generated humans images. The research particularly focuses on the effects of facial behavior display, including facial expressions, head movements, and eye movements, on audience perception.

Using a CG animated human head, the study examines how these behaviors influence believability, eeriness, and accurate behavior recognition among participants. The findings indicate that facial movements impact the perceived eeriness of CG humans, while the combination of facial expressions and head movements enhances believability. Additionally, the accuracy of behavior recognition improves with the inclusion of head movements. This study underscores the importance of detailed and realistic facial behaviors in achieving lifelike and believable CG human characters, which is important for applications in virtual reality, gaming, and animation [Qiao and Eglin 2011].

Zhou et al. [Zhou and Kawabata 2023] conducted study that explored into the subjective realms of beauty, liking, valence, and arousal by offering insights into emotional responses to AI-generated images. Their study utilized a diverse set of images and involved participants in rating their emotional reactions to each image. The findings revealed the complex interplay between visual aesthetics and emotional responses, providing a deeper understanding of how generative models can be tuned to evoke specific emotions. This research is particularly relevant for applications in advertising, social media, and entertainment, where emotional engagement is a key factor [Zhou and Kawabata 2023].

Treder et al. [Treder et al. 2022] focused on the subjective evaluation of image quality through a human-centered study. Their research prioritized human ratings to gauge various dimensions of image quality, including sharpness, color fidelity, and the presence of artifacts. By comparing human assessments with computational metrics, the study highlighted the discrepancies that often arise between human perception and algorithmic evaluations. This work underscored the need for more reliable and perceptually aligned metrics to accurately assess the quality of AI-generated images, which is critical for applications in fields such as digital art and photography [Treder et al. 2022].

Sarkar et al. [Sarkar et al. 2021] introduced HumanGAN, a study demonstrating the capabilities of generative models in crafting lifelike images. Their research focused on evaluating the realism of images generated by GANs through human assessments. Participants were asked to distinguish between real and AI-generated images, providing insights into the effectiveness of GANs in producing photorealistic content. This study highlighted the potential of GANs in various domains, including entertainment and virtual reality, where the ability to create lifelike images is highly valued [Sarkar et al. 2021].

Ha et al. [Ha et al. 2024] conducted a comprehensive study examining the ability to distinguish between human art and AI-generated images and discussed the challenges posed by the growing prevalence of generative AI in the art world. They utilized various approaches, including classifiers trained via supervised learning, targeting diffusion models, and the expertise of professional artists. The authors curated a diverse collection of real human art across seven distinct styles and generated corresponding images using five different generative models. They then employed eight different detection methods, encompassing both computational detectors and human evaluators, including 180 crowdworkers, over 4000 professional artists, and 13 expert artists with experience in identifying AI-generated images. The study found that while automated detectors like Hive and expert artists performed well, each had unique vulnerabilities: Hive struggled against adversarial perturbations, and expert artists had higher false positive rates. The findings underscore the importance of a combined approach, leveraging both human and automated detectors, to achieve optimal accuracy and robustness in distinguishing AI-generated images from human art [Ha et al. 2024].

Rassin et al. [Rassin et al. 2022] provided an in-depth analysis of text-image alignment within the DALL-E 2 framework. Their research examined how well the generated images matched the given textual descriptions, involving human participants in evaluating the accuracy and coherence of these alignments. The study revealed the strengths and limitations of DALL-E 2 in generating contextually appropriate images, emphasizing the importance of precise text-image alignment in applications like automated content creation and digital marketing [Rassin et al. 2022].

Hulzebosch et al. addressed the challenge of detecting Convolutional Neural Network-generated facial images in real-world scenarios by proposing a comprehensive evaluation framework. This framework includes cross-model, cross-data, and post-processing evaluations to assess the robustness of detection methods. The study revealed that although various algorithmic detection methods exist, they often generalize poorly to data from unknown sources, rendering them less effective in practical applications. Additionally, the research evaluated the effectiveness of common image pre-processing techniques and conducted an online survey to assess human performance in detecting CNN-generated images. The findings indicate that current CNN-based detection methods lack the robustness required for reliable use in real-world scenarios, highlighting the need for improved techniques [Hulzebosch et al. 2020].

Degardin et al. [Degardin et al. 2024] conducted research on generative images depicting human actions. Their study focused on the accuracy and realism of action representations in AI-generated images, involving human evaluations to assess the fidelity of these depictions. The findings highlighted the challenges in generating complex action scenes and the importance of human feedback in refining these models. This research is particularly relevant for applications in gaming and animation, where realistic action sequences are important [Degardin et al. 2024].

Sun et al. [Sun et al. 2024] explored gender bias in AI-generated images within the DALL-E 2 framework. Their study involved analyzing the representation of different genders in generated images and assessing the presence of any biases. Human participants were involved in evaluating the fairness and accuracy of these representations. The research highlighted the ethical considerations in AI-generated content and the need for unbiased generative models. This study is significant for applications in media and advertising, where fair representation is essential [Sun et al. 2024].

Xu et al. conducted an extensive study on the influence of AI-generated content on text-image retrieval models, which is highlighting an invisible relevance bias introduced by AI-generated images. Their research involved constructing a benchmark to explore this bias and performing extensive experiments that demonstrated how retrieval models tend to rank AI-generated images higher than real images, despite the AI-generated images not being more visually relevant to the queries. This bias was found to be prevalent across various retrieval models, irrespective of their training data and architectures. Furthermore, the inclusion of AI-generated images in the training data exacerbated this bias, creating a vicious cycle that intensified the relevance bias. To address this issue, Xu et al. proposed an effective training method aimed at debiasing the models. Their method revealed that AI-generated images cause the image encoder to embed additional information, leading to higher relevance scores for these images. This study underscores the need for robust debiasing methods in the development of text-image retrieval models to ensure fair and accurate retrieval outcomes [Xu et al. 2023].

These studies collectively underscore the critical role of human evaluations in assessing the quality, realism, and alignment of AI-generated images. Despite significant advancements in computational metrics, there remains an important opportunity to design and develop a statistically validated comprehensive study that addresses the key dimensions of photorealism, image quality, and text-image alignment simultaneously. The objective is to develop a questionnaire that closely aligns with the workings of computational image metrics, facilitating a fair assessment of both images and metrics.

## 2.4 Scaling Strategies

Scaling strategies play a critical role in image quality assessment by scaling metric scores with human perceptual judgments for comparison. Traditional methods such as Z-standardization and Min-Max scaling can introduce biases, often disproportionately emphasizing certain scores over others, such as giving undue weight to Peak Signal-to-Noise Ratio (PSNR) at the expense of Structural Similarity Index Measure (SSIM). To mitigate these biases, advanced non-linear methods have been developed to align metric scores more closely with mean opinion scores (MOS) from human evaluators, providing a more accurate evaluation of image quality [Rehman et al. 2015].

Advanced models, such as the Non-linear Receptive Field model dynamically optimizing parameters to enhance correlation with human ratings, tailored to specific image datasets [Luna et al. 2023]. This model adjusts its parameters based on the visual complexity of the dataset, thus providing a more precise mapping between computed metric values and human perceptual scores. The Non-linear Receptive Field model has demonstrated improved performance in aligning with human judgments by particularly in datasets featuring complex visual features.

Table 2. Visual Verity: Demography Questionnaire

| Demography Questionnaire | |
| --- | --- |
| **Question ID** | **Question Text** |
| DQ1G | What is your Gender? |
| DQ2G | What is your age? |
| DQ3G | What is your Educational Qualification? |
| DQ4G | Experience with AI or Computer-Generated Images. |
| DQ5G | How often do you view digital images or graphics (including online, in games, movies, etc.)? |
| DQ6G | Do you have experience in graphic design, photography, or a related field? |
| DQ7G | What is your country of residence? |

Moreover, the Saliency-Guided Local Full-Reference Image Quality Assessment leverages visual saliency to weight local image quality scores, prioritizing regions likely to capture human attention. This approach is based on the premise that not all image regions contribute equally to perceived quality; therefore, it focuses on areas that are more salient to the human visual system [Varga 2022]. By integrating saliency maps into the assessment process, this method aims to provide a more accurate reflection of human visual perception, improving the correlation with subjective quality ratings.

Another significant scaling strategy is the use of Multi-Task Learning (MTL) frameworks, which can predict human perceptual scores by training on multiple related tasks simultaneously. MTL frameworks help leverage shared representations across different tasks, thereby improving the generalization and robustness of the quality assessment models [Huang et al. 2022]. These frameworks have been particularly effective in handling diverse and heterogeneous datasets, offering a scalable solution to image quality assessment.

Deep learning-based models for scaling, such as those incorporating convolutional neural networks (CNNs) and transformer architectures, have gained prominence for their ability to learn feature representations from large datasets and align them with human perceptual scores through end-to-end training [Zhang et al. 2018]. By utilizing large-scale annotated datasets, these models can capture complex patterns and nuances in visual data, making them highly effective for quality assessment.

Despite the advancements brought by these methods, their complexity and black-box nature often render them less transparent and harder to trust in practical applications. This underscores a critical gap in the field: the need for an interpretable and fair scaling strategy that practitioners can readily trust. The development of transparent scaling methods addresses this gap by providing robust, interpretable, and fair solutions for scaling image quality metrics, thereby enhancing their practical applicability and acceptance.

In response to these challenges, we already proposed the Interpolative Binning Scale (IBS) [Aziz et al. 2024], which is an innovative scaling strategy designed to improve the interpretability and fairness of quality assessments. Unlike traditional methods, IBS employs a binning approach to segment metric scores into discrete intervals, which are then interpolated to provide a continuous and smooth scaling. This method ensures that the scaling is not biased by extreme values and provides a more balanced representation of the metric scores across the entire range. By aligning the interpolated scores with human perceptual data, IBS offers a transparent and easily interpretable scaling method that can be trusted by practitioners. In this paper, we present an expert evaluation of IBS to demonstrate its effectiveness in aligning computational metric scores with human visual perception [Aziz et al. 2024].

## 3 STUDY 1: DEVELOPMENT AND VALIDATION OF A QUESTIONNAIRE ON PHOTOREALISM, IMAGE QUALITY, AND TEXT-IMAGE ALIGNMENT

This section provides detailed information on the development and statistical validation of a questionnaire designed to evaluate photorealism, image quality, and the alignment of text with images. The proposed questionnaire, named Visual Verity, is structured into four components: Demography, Photorealism, Image Quality, and Text-Image Alignment. Each component is designed with a specific rationale to ensure comprehensive evaluation and accurate data collection.

Visual Verity received ethics approval from the Non-Medical Research Ethics Board (NMREB) at the University of Western Ontario (UWO), Canada, under project ID: 124753. The official approval is publicly accessible at: https://github.com/udanish50/VisualVerity/.

### 3.1 Visual Verity: Demography Questionnaire

The Demography Questionnaire (Table 2) is designed to collect essential demographic data from participants, which is important for understanding the diversity within our sample. We collect limited demographic information to ensure participant privacy, as our questionnaire is partially anonymous.

Table 3. Visual Verity: Photorealism Questionnaire

| Photorealism | |
|---|---|
| **Question ID** | **Question Text** |
| PQ1R | The image looks like a photograph of a real scene. |
| PQ2R | I can easily imagine seeing this image in the real world. |
| PQ3R | The visual details in this image make it appear realistic. |
| PQ4R | The textures in the image look natural and real. |
| PQ5R | The lighting and shadows in the image contribute to its realism. |

The questionnaire begins with DQ1G, which asks about gender to identify any potential gender bias in assessing AI-generated images. Gender diversity is important for analyzing differences in perception and evaluation, as previous studies [Zhou and Kawabata 2023] have shown gender can influence how visual information is processed and interpreted. DQ2G inquires about the participant's age by recognizing that familiarity with technology and visual media often varies across age groups. Olson et al. [Olson et al. 2011] reported that younger individuals, such as those aged 18–29, tend to use and understand technology more extensively than older adults, which is impacting their interactions with AI-generated content [Olson et al. 2011].

DQ3G asked about Educational background, as it correlates with critical thinking and analytical skills. Individuals with higher educational attainment, such as master's or doctoral degrees, are generally expected to show more advanced critical thinking abilities compared to those with less education. This question helps us understand the educational diversity of our participants and its potential impact on image evaluation. DQ4G seeks to assess the level of experience with AI or computer-generated images. This question is particularly important for identifying biases related to familiarity with AI technology. By asking whether participants have prior experience with AI-generated images, we can better interpret their evaluations and ensure a more balanced assessment when linked with other demographic factors such as gender.

The frequency of interaction with digital media is explored in DQ5G, which asks participants how often they view digital images or graphics, including online content, games, and movies. Regular exposure to digital media can enhance one's ability to evaluate image quality and realism. DQ6G questions the participant's hands-on experience in creative fields like graphic design or photography. Such experience is likely to provide participants with a more discerning eye for evaluating image quality and artistic elements. The last question DQ7G asks about the country of residence to account for cultural and regional differences that might affect the interpretation of images. Cultural and regional backgrounds can influence visual perception by making it essential to consider geographical diversity in our study.

These questions collectively aim to establish a comprehensive profile of the respondents by enabling analysis of how demographic factors influence the reception and evaluation of AI-generated images. Table 10 presents the overall respondent profiles of our study.

## 3.2 Visual Verity: Photorealism Questionnaire

The Photorealism Questionnaire (Table 3) is designed to evaluate the perceived realism of AI-generated images. It will help to understand how photorealism is important for applications ranging from virtual reality to digital art, where the believability of an image can significantly impact user experience. This section provides the rationale behind each question included in the questionnaire. The Likert scale with the following options: strongly agree, somewhat agree, neutral, somewhat disagree, and strongly disagree is used in all questions.

The questionnaire begins with PQ1R, which asks if the image looks like a photograph of a real scene. This question is fundamental as it directly probes the participant's immediate perception of realism, which is a primary indicator of photorealism in AI-generated images. The second question PQ2R follows by asking whether participants can easily imagine seeing the image in the real world. This question addresses the contextual plausibility of the image, an important factor in photorealism [Shadiev et al. 2020].

The third question, PQ3R, focuses on the visual details that contribute to the image's realism. High levels of detail, such as sharpness and clarity, are key components of photorealistic images. The detailed images are more likely to be perceived as realistic because they mimic the high-resolution perception of human vision. PQ4R asks about the natural appearance of textures in the image. Texture realism is a critical aspect of photorealism, as realistic textures can significantly enhance the believability of an image. The accurate texture representation is important for the realistic depiction of surfaces and materials in synthetic images.

Finally, PQ5R evaluates the contribution of lighting and shadows to the image's realism. Proper lighting and shadowing are essential for creating depth and dimensionality, which are pivotal for photorealism. The lighting effects can dramatically influence the perception of realism in digital images. These questions collectively aim to provide a comprehensive assessment of the photorealism of AI-generated images, allowing for nuanced insights into how different aspects of an image contribute to its overall realism.

## 3.3 Visual Verity: Image Quality Questionnaire

The Image Quality Questionnaire (Table 4) is designed to evaluate the perceived quality of AI-generated images. Assessing image quality is important for numerous applications, including digital media, virtual reality, and e-commerce, where clarity, color accuracy, and the absence of distortions significantly impact user experience and satisfaction. This section details the rationale behind each question included in the questionnaire. The Likert scale with the following options: strongly agree, somewhat agree, neutral, somewhat disagree, and strongly disagree, is utilized in all questions.

The questionnaire begins with IQ1G, which asks if the image is clear and sharp. Image clarity and sharpness are fundamental aspects of quality, directly influencing how detailed and lifelike an image appears. The second question IQ2G is asking about the vibrancy and lifelikeness of colors in the image. The color accuracy and vibrancy are critical for creating visually appealing images that accurately represent the intended scene or subject. The third question, IQ3G, seeks participants' overall satisfaction with the image quality. This question provides a holistic measure of perceived image quality by capturing participants' general impressions and satisfaction levels. IQ4G asks about the presence of visible artifacts or distortions in the image. Artifacts and distortions can significantly degrade the perceived quality of an image, affecting its realism and usability. Finally, IQ5G evaluates whether the resolution of the image meets participants' expectations. These questions collectively aim to provide a comprehensive assessment of the image quality of AI-generated images.

## 3.4 Visual Verity: Text-Image Alignment Questionnaire

The Text-Image Alignment Questionnaire (Table 5) is designed to evaluate the degree to which AI-generated images align with their corresponding textual descriptions. It has applications in automated content creation, digital marketing, and educational tools, where the coherence between text and visuals significantly impacts user engagement and comprehension. Each question in the questionnaire is designed to assess different facets of alignment by ensuring a comprehensive evaluation of this important aspect. The Likert scale with the following options: strongly agree, somewhat agree, neutral, somewhat disagree, and strongly disagree is used in all questions.

The questionnaire begins with CQ1M, which asks whether the image perfectly aligns with the given caption. This question addresses the overall coherence between the text and the visual content. Perfect alignment is essential for ensuring that the generated image accurately represents the described scene, which is particularly important in fields like advertising and educational content. CQ2M follows, asking if the elements in the image correspond to the described scene in the caption. This question focuses on the specific components and details within the image, ensuring that all elements mentioned in the caption are accurately depicted. The third question, CQ3M, seeks to understand if participants would describe the image with a caption that closely matches the provided one. This question evaluates the naturalness and intuitiveness of the text-image pair, ensuring that the generated captions feel authentic and accurate to human observers.

CQ4M asks if the image misses some details mentioned in the caption. This question is critical for identifying gaps or omissions in the generated image, which can detract from the perceived accuracy and completeness of the representation. Ensuring that all details are correctly depicted is important for maintaining the credibility and reliability of AI-generated content. The last question CQ5M, evaluates whether the image is perceived as a true representation of the given caption. This question provides a holistic measure of the text-image alignment, capturing the overall impression of accuracy and coherence. A strong alignment between text and image is vital for applications where users rely on visual content to understand or complement textual information.

These questions collectively aim to provide a thorough assessment of the alignment between text and AI-generated images, enabling a nuanced analysis of how well these images meet the expectations set by their descriptions.

## 3.5 Questionnaire Validation

The subsection will provide details on scale refinement and validation by exploratory factor analysis, confirmatory factor analysis using partial least squares and structural equation modelling.

### 3.5.1 Scale Refinement and Validation by Exploratory Factor Analysis on, N=200). 
We carried out an Exploratory Factor Analysis (EFA) on the first group of 200 participants. This method helped us uncover the basic factor pattern of the questions in our survey and to check that each question accurately matched the specific category it was meant to measure. We found that the data was a good fit for this kind of analysis because the Kaiser-Meyer-Olkin Measure of Sampling Adequacy was 0.906. This high number means our data was suitable for the analysis. Additionally, Bartlett's Test of Sphericity gave us a Chi-Square value of around 2141, which confirmed that our questions were interrelated enough to proceed with a valid factor analysis.

The results of the EFA clearly showed three separate categories, which we had intended to measure: Photorealism, Image Quality, and text-image alignment. The analysis told us how many underlying factors there were, how strongly each survey question was connected to its factor, and how much of the overall variation was explained by these factors. This confirmed that our questionnaire was well-constructed and could effectively measure the different aspects we were interested in, as shown in the referenced table 6.

### 3.5.2 Confirmatory Factor Analysis (CFA) using Partial Least Squares (PLS) Structural Equation Modeling (SEM) (N=300). 
WarpPLS 8.0 was employed to perform scale validation for the constructs of Photorealism, Image Quality, and Text-Image Alignment using the study's two Paper (i.e., 300). CFA was conducted to confirm the factor structure identified in the EFA by providing a more rigorous validation of the survey's construct validity. This approach is suited for the assessment of the measurement model (i.e., based on reflective models comprising Photorealism, Image Quality, and Text-Image Alignment) as it can estimate the scale validation without developing the structural model [Abdul-Latif and Abdul-Talib 2017].

Since the study's scale is based on reflective constructs, the estimation of reflective constructs includes the assessment of outer

Table 4. Visual Verity: Image Quality Questionnaire

| Image Quality | |
|---|---|
| **Question ID** | **Question Text** |
| IQ1G | The image is clear and sharp. |
| IQ2G | The colors in the image are vibrant and lifelike. |
| IQ3G | I am satisfied with the overall quality of this image. |
| IQ4G | The image has no visible artifacts or distortions. |
| IQ5G | The resolution of the image meet my expectations. |

Table 5. Visual Verity: Text-Image Alignment

| Text-Image Alignment | |
|---|---|
| **Question ID** | **Question Text** |
| CQ1M | The image perfectly aligns with the given caption. |
| CQ2M | The elements in the image correspond to the described scene in the caption. |
| CQ3M | If I were to describe this image with a caption, it would closely match the provided one. |
| CQ4M | The image misses some details mentioned in the caption. |
| CQ5M | I feel the image is a true representation of the given caption. |

Table 6. EFA (Study 1 - Comprising Camera Generated and DALLE Images)

| Constructs | Items | Loading Dim-1 | Loading Dim-2 | Loading Dim-3 |
|---|---|---|---|---|
| Photorealism (Eigen Value: 7.276, Cronbach's Alpha: 0.923) | PQ1R | 0.913 | 0.533 | 0.553 |
| | PQ2R | 0.861 | 0.49 | 0.481 |
| | PQ3R | 0.915 | 0.536 | 0.579 |
| | PQ4R | 0.872 | 0.507 | 0.677 |
| | PQ5R | 0.799 | 0.38 | 0.636 |
| Text-Image Alignment (Eigen Value: 1.787, Cronbach's Alpha: 0.925) | CQ1M | 0.487 | 0.925 | 0.402 |
| | CQ2M | 0.547 | 0.889 | 0.463 |
| | CQ3M | 0.489 | 0.879 | 0.31 |
| | CQ5M | 0.512 | 0.916 | 0.466 |
| Image Quality (Eigen Value: 1.077, Cronbach's Alpha: 0.884) | IQ1G | 0.548 | 0.361 | 0.867 |
| | IQ2G | 0.486 | 0.314 | 0.84 |
| | IQ3G | 0.652 | 0.455 | 0.871 |
| | IQ5G | 0.566 | 0.456 | 0.855 |

*Note: CQ4M and Image IQ4G are deleted due to cross-loadings*

loadings, convergent validity (AVE), and reliabilities. Hair et al. [Hair and Alamer 2022] suggested that the outer loading should be 0.60 or greater, convergent validity (i.e., Average Variance Extracted – AVE) ought to be 0.50 or above, and reliabilities should exceed the value 0.70. Table 2 shows that each item's loading on its respective factor was critically evaluated. Item loadings for each scale, ranging from 0.833 to 0.944, which indicate strong individual contributions to their respective constructs. Cronbach's Alpha for Photorealism, Image Quality, and Text-Image Alignment were 0.931, 0.952, and

0.94, respectively, suggesting high internal consistency. Composite Reliability ranged from 0.948 to 0.957, and AVE values from 0.784 to 0.847, affirming the convergent validity of the scales, as shown in Table: 6.

This step was important to ensure that each of the three scales measured distinct constructs and was not overly inter-correlated. Discriminant validity was assessed using the HTMT ratio [Henseler et al. 2015]. This method involved comparing the correlations between constructs against the correlations within constructs. The HTMT ratios for constructs such as Text-Image Alignment and Image Quality were 0.526 and 0.65, respectively as shown in Table 8. These values being below the threshold of 0.90 provided strong evidence of discriminant validity among the scales, as shown in Table: 8.

Table 7. Confirmatory Analysis (300 Sample based on other three Images GI)

| Construct | Items | Loads | Type | SE | P value | Cronbach Alpha | Comp. Reliability | AVE | FVIF |
|-----------|-------|-------|------|-----|---------|----------------|-------------------|-----|------|
| Photorealism | PQ1R | 0.909 | Reflect | 0.05 | < 0.001 | 0.931 | 0.948 | 0.784 | 2.094 |
| | PQ2R | 0.833 | Reflect | 0.05 | < 0.001 | | | | |
| | PQ3R | 0.909 | Reflect | 0.05 | < 0.001 | | | | |
| | PQ4R | 0.884 | Reflect | 0.05 | < 0.001 | | | | |
| | PQ5R | 0.888 | Reflect | 0.05 | < 0.001 | | | | |
| Text-Image Align. | CQ1M | 0.944 | Reflect | 0.05 | < 0.001 | 0.952 | 0.965 | 0.874 | 1.623 |
| | CQ2M | 0.927 | Reflect | 0.05 | < 0.001 | | | | |
| | CQ3M | 0.932 | Reflect | 0.05 | < 0.001 | | | | |
| | CQ5M | 0.936 | Reflect | 0.05 | < 0.001 | | | | |
| Image Quality | IQ1G | 0.937 | Reflect | 0.05 | < 0.001 | 0.94 | 0.957 | 0.847 | 2.547 |
| | IQ2G | 0.895 | Reflect | 0.05 | < 0.001 | | | | |
| | IQ3G | 0.922 | Reflect | 0.05 | < 0.001 | | | | |
| | IQ5G | 0.926 | Reflect | 0.05 | < 0.001 | | | | |

The comprehensive EFA, CFA, and discriminant validity assessments provided robust evidence of the survey's validity and reliability. The survey effectively measures the constructs related to AI-generated images, specifically Photorealism, Image Quality, and Text-Image Alignment, making it a valuable tool in this field.

### 3.6 Discussion and Implications

The process of developing and validating the Visual Verity questionnaire highlights several key considerations and implications for the evaluation of AI-generated images. The demography ensures that participant diversity is well-documented by providing critical insights into how demographic factors like gender, age, educational background, and cultural context impact the evaluation of AI-generated images. This demographic profiling is essential for identifying biases and variations in perception, which can inform more equitable and inclusive AI development practices.

The photorealism component of the questionnaire addresses the immediate perceptual aspects of AI-generated images, such as their resemblance to real-world scenes and the natural appearance of textures and lighting. By focusing on these elements, the questionnaire ensures that the generated images meet high standards of realism, which is crucial for applications in virtual reality, digital art, and gaming. In assessing image quality, the questionnaire evaluates clarity, color vibrancy, and the absence of distortions. High image quality is critical for user satisfaction in digital media, e-commerce, and other visual-centric applications. By asking participants to rate these aspects, the questionnaire helps identify areas where AI-generated images may fall short. The text-image alignment section focuses on the coherence between textual descriptions and generated images, which is vital for content creation, marketing, and educational tools. Ensuring that images accurately reflect their descriptions enhances the utility and effectiveness of AI systems in these domains. The detailed questions about alignment help pinpoint discrepancies between text and image.

The validation processes, including Exploratory Factor Analysis (EFA) and Confirmatory Factor Analysis (CFA), confirm that the Visual Verity questionnaire is a robust tool for evaluating AI-generated images. These statistical validations ensure that each section of the questionnaire accurately measures its intended constructs by providing reliable data for analysis. The rigorous validation process enhances the credibility and applicability of the questionnaire in diverse research and development settings.

The findings from the validation and application of the Visual Verity questionnaire have several implications. First, they highlight the need for comprehensive evaluation tools that consider multiple dimensions of image perception. As result of validation we excluded one item from image quality and text-image alignment. Second, they underscore the importance of demographic diversity in understanding how different groups perceive AI-generated images. They point to the potential for using detailed feedback to improve the realism, quality, and contextual relevance of AI-generated images, thereby enhancing their applicability across various domains.

## 4 STUDY 2: BENCHMARKING HUMAN PERCEPTION FOR PHOTOREALISM, IMAGE QUALITY, AND TEXT-IMAGE ALIGNMENT AGAINST CAMERA-GENERATED AND AI-GENERATED IMAGES

This section outlines the details about datasets, preprocessing and presentation of images, analytical and statistical analysis, and statistical validation of each category and associated discussion.

Table 8. Discriminant Validity using HTMT Ratios

|  | Photorealism | Image Quality | Text-Image Alignment |
| --- | --- | --- | --- |
| **Photorealism** | - | - | - |
| **Image Quality** | 0.771 | 0.65 | - |
| **Text-Image Alignment** | 0.526 | - | - |

*Note: HTMT ratios: (good if < 0.90, best if < 0.85)*

## 4.1 Materials

Our study utilized the MS COCO (Microsoft Common Objects in Context) dataset, a foundational resource and large-scale dataset consisting of 328K images [Lin et al. 2014b]. The MS COCO dataset is renowned for its extensive collection of images, each annotated with distinct object categories and accompanied by human-written captions. These captions provide a diverse array of descriptions, capturing the essence and context of the scenes depicted. To ensure a manageable and focused study, we selected a subset of 20 images from the MS COCO dataset. This selection process was driven by practical considerations, as conducting a survey on images with human participants typically requires 20 to 30 minutes per participant. By choosing a diverse range of categories within this subset, we ensured that our study covered a broad spectrum of scenes and contexts. Each image in this subset was paired with its corresponding human-written caption.

We generated corresponding images using four distinct AI generative models: DALL-E2, DALL-E3, Stable Diffusion, and Glide. Each model received the same human-written caption associated with the selected images by allowing us to compare the performance of different AI technologies and understand the architectural differences among these models [Aziz et al. 2024]. The image generation process was carried out using the NVIDIA Corporation GA102GL [RTX A6000] (rev a1).

The architecture of each model utilizes different technologies. DALL-E2 and DALL-E3 leverage a combination of transformer architectures, with DALL-E2 incorporating CLIP embeddings to enhance its ability to generate high-resolution, photorealistic images with improved contextual understanding. Stable Diffusion operates by iteratively refining Gaussian noise into detailed images, which is notable for its high-resolution output and flexibility. GLIDE uses the principles of diffusion models by conditioning the image generation process on textual information and is known for upscale images while maintaining high photorealism. Due to the industry-oriented nature of both DALL E and Stable Difusion, only high-level architectural detail is available, which allows us to conduct on a diverse range of models. By employing these diverse AI generative models and utilizing state-of-the-art hardware, our study aims to benchmark human perception of photorealism, image quality, and text-image alignment.

To ensure a fair comparison, we reshaped all images to a uniform size of 299x299 pixels. This resizing was accomplished using linear interpolation [Lin et al. 2014a]. Linear interpolation is a technique known for preserving the original image quality while adjusting dimensions [Aziz et al. 2024]. This reshaping of images is important to ensure that all images, irrespective of their source or original dimensions were presented for evaluation in a fair environment. This is also important and required by computational image metrics assessments because these metrics require images should be of the same shape.

## 4.2 Recruitment of Study Participants

We recruited 350 study participants from Prolific [Prolific 2024], an online platform recognized for its diverse and highly relevant participant pool. This platform ensured that our study engaged individuals capable of providing valuable insights on the images evaluated. To deliver our survey, we utilized the Qualtrics platform, known for its robust and user-friendly interface [Qualtrics 2024].

To ensure thoughtful completion of the survey, we compensated participants at a rate of $10.28 CAD per hour through the Prolific platform. In accordance with guidelines provided by the Non-Medical Research Ethics Board (NMREB), participants were compensated even if they chose to leave the study before completion, emphasizing our commitment to ethical standards.

Participants were presented with camera and AI-generated images accompanied by a caption, with the images displayed in a randomized order to prevent any order bias. The origin of the image was not disclosed. This was important to ensure unbiased assessments based solely on the visual and contextual content of the images, free from any preconceived notions about the AI models involved.

To encourage thorough evaluation, participants were required to spend a minimum of 120 seconds assessing each image. Each participant evaluated one camera-generated image and four AI-generated images, ensuring a comprehensive assessment across different types of imagery. According to Prolific's statistics, participants spent an average of 16 minutes and 5 seconds completing the survey, which indicates substantial engagement with the task. For the image assessments, we employed a Likert scale ranging from 'Strongly Disagree' to 'Strongly Agree', allowing us to capture nuanced evaluations of the images' quality, realism, and text-image alignment.

## 4.3 Objective Properties of Images

We analyzed the objective properties of camera-generated and AI-generated images to identify significant differences in their key characteristics (see Table 9). This analysis involved calculating metrics such as hue, saturation, brightness, color vibrancy, and entropy, which provide a quantitative basis for distinguishing between the Camera and AI-generated images [Zhou and Kawabata 2023].

Table 9. Mean Values of Objective Properties by Images Generated by Camera and Different Generative Models

| Model | Hue | Saturation | Brightness | Vibrancy | Entropy |
|---|---|---|---|---|---|
| Camera Generated | 49.0062 | 72.0073 | 131.7686 | 201.9793 | 7.1148 |
| DALL-E 2 | 52.0629 | 87.0332 | 149.7148 | 224.2862 | 7.4418 |
| DALL-E 3 | 52.5353 | 89.5278 | 136.9140 | 205.6381 | 7.6429 |
| GLIDE | 64.3703 | 85.7061 | 144.1945 | 212.4636 | 6.9872 |
| Stable Diffusion | 59.5445 | 65.4708 | 136.0586 | 209.4821 | 7.5189 |

Hue represents the dominant color wavelength in an image. Our analysis revealed a clear distinction between camera-generated and AI-generated images. Camera-generated images had a mean hue value of 49.0, whereas AI-generated images had hue values exceeding 52. This indicates a notable difference, which suggests that the hue in camera-generated images tends to be more varied and potentially more naturally distributed compared to AI-generated ones.

Saturation measures the intensity of the color. Camera-generated images had a mean saturation value of 72.0073, which is lower than that of AI-generated images, where the values ranged from 65.4708 to 89.5278. This difference shows that AI-generated images tend to have more vivid and intense colors, possibly due to the algorithms' tendencies to enhance colors to make the images appear more visually striking.

Brightness refers to the lightness or darkness of an image. The mean brightness of camera-generated images was 131.76, while AI-generated images had higher brightness levels, with DALL-E 2 achieving the highest mean brightness of 149.71. This difference suggests that AI-generated images may be brighter, likely a result of generative models' design to create visually appealing and well-lit images.

Color vibrancy quantifies the richness of colors in an image. Camera-generated images had a mean vibrancy of 201.98, whereas AI-generated images showed higher vibrancy, with DALL-E 2 reaching 224.28. This indicates that AI-generated images often have richer and more dynamic color palettes, which can make them appear more engaging and aesthetically pleasing.

Entropy measures the complexity and amount of information in an image. Camera-generated images had a mean entropy of 7.11, while AI-generated images ranged from 6.98 to 7.64. Higher entropy values in AI-generated images, which is particularly in DALL-E 3 and DALL-E 2, suggest that these images might contain more visual details and complexity, possibly due to the sophisticated algorithms used in their generation.

These objective properties highlight the fundamental differences between camera-generated and AI-generated images. Understanding these differences paves the way for developing metrics to distinguish between the two. The figure 2 and table provide a visual and tabular representation of these findings.

### 4.4 Respondent Profiles and Analysis

The demographics of our respondents are presented in Table 10. For DQ1G, where the gender of participants was asked, 57% identified as male, 42% as female, and only 1% as transgender or non-binary. This is a reflection of global population statistics, where the transgender or non-binary community is less than 1% [World Papulation Review 2024a]. This suggests that our respondent pool is diverse, although there is a higher percentage of men than women. When linked to DQ4G, which asked about experience with AI imaging, it was found that 55% of the women had familiarity with generative AI, compared to 50% of the men. It indicates a significant engagement of women in AI-related fields.

In DQ3G, which asked about the education level, 48% of the respondents had an undergraduate or bachelor's degree, 13% had a master's degree, and only 2% had a doctoral degree. This distribution aligns with global statistics, where only a small percentage of the population holds a PhD [World Papulation Review 2024b]. The high percentage of well-educated respondents ensures thoughtful and informed responses.

For DQ2G, the age distribution was diverse, with the largest groups being 18-24 years and 25-34 years, each constituting 38% of the respondents. This suggests that our survey primarily engaged young adults, who are generally more familiar with technology and digital media. The remaining age groups were 35-44 years (14%), 45-54 years (5%), 55-64 years (2%), and 65-74 years (1%), providing a broad spectrum of perspectives.

In DQ4G, regarding familiarity with AI or computer-generated images, 49% of the respondents were familiar but did not have hands-on experience, while 28% had some experience through courses or hobbies. Only 4% were experts working professionally with AI or CGI, 18% had heard of AI but did not know much about it, and 2% were not familiar at all. This range of familiarity levels indicates a well-rounded respondent pool with varied exposure to AI technologies.

DQ5G inquired about the frequency of viewing digital images or graphics. Nearly half of the respondents (48%) viewed digital images multiple times a day, 30% did so daily, 17% weekly, 4% monthly, and 2% rarely. This high frequency of interaction with digital media suggests that the respondents are well-positioned to evaluate the quality and realism of the images presented in the study.

For DQ6G, which asked about hands-on experience in creative fields like graphic design or photography, 45% of the respondents appreciated the arts and often viewed related content, but did not have direct experience. Another 28% engaged in these activities as a hobby or pastime, while 9% had taken courses or received training, and 7% were professionals in the field. This distribution highlights a substantial appreciation and understanding of visual arts among the respondents.

DQ7G captured the geographical diversity of the respondents, with significant representation from countries like the Republic of South Africa (14%), the Portuguese Republic (13%), and Poland (13%). Other countries included Pakistan (11%), the United Kingdom (8%), the United Mexican States (6%), Hungary (3%), the Hellenic Republic (Greece) (4%), the Kingdom of Spain (3%), the Republic of Italy (4%), and the United States of America (5%). This international representation ensures that the study benefits from a wide range of cultural perspectives, enriching the analysis.

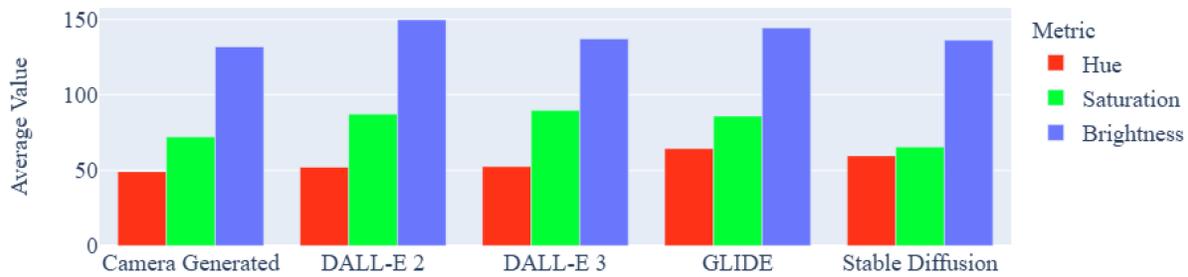

Fig. 2. Mean Values of Objective Properties such as Hue, Saturation and Brightness of Camera and Generated Images

The responses from less than 2% of participants came from France, Estonia, Slovenia, Czech Republic, Canada, Belgium, Zimbabwe, Germany, India, Portugal, and Scotland, among others. This broad geographic distribution underscores the global relevance and inclusiveness of our study, contributing to a comprehensive understanding of how different demographic factors influence the reception and evaluation of AI-generated images.

### 4.5 Statistical Analysis: Photorealism

The data presented in Table: 11 provides a comprehensive overview of participant responses, with mean scores (out of 5) across camera and AI-generated images, while Table (12) showing question wise responses. The Camera-generated images received the highest mean score of 4.06, which indicates that participants found these images to be the most realistic, which is quite obvious. DALL-E2 followed with a score of 3.63, which suggests that it produces relatively realistic images, although not quite on par with camera-generated ones. GLIDE, on the other hand, scored lower at 2.04, which indicates that its images are perceived as less realistic. Stable Diffusion and DALL-E3 scored 3.30 and 2.75 respectively by placing them between the extremes set by camera-generated images and GLIDE. This spread shows the varying capabilities of different AI models in achieving photorealism, with camera-generated images still setting the benchmark.

We further analyzed deeper questions (See Table: 12), For the PQ1R (the image looks like a photograph of a real scene), camera-generated images received the highest "Strongly Agree" responses (49%), followed by DALL-E2 (39%). GLIDE had the lowest (10%) and the highest "Strongly Disagree" rate (53%), indicating it was perceived as the least realistic. Similarly, In PQ2R, camera-generated images were most contextually plausible (56% "Strongly Agree"), with GLIDE again scoring the lowest (9% "Strongly Agree"). For visual details (PQ3R), textures (PQ4R), and lighting (PQ5R), camera-generated images consistently ranked highest, while GLIDE scored the lowest, showing shortcomings in achieving realism.

Before proceeding with further statistical analysis, our research question was formulated as follows: Our aim is to identify disparities in photorealism among various AI models compared to camera-generated images. Our hypothesis aims to explore the disparities in perceived photorealism among different types of images, including those generated by cameras and various AI models such as DALL-E2, DALL-E3, Stable Diffusion, and GLIDE. The objective is to ascertain the extent and nature of differences in photorealism across these distinct image sources.

We detail the independent and dependent variables used in our study to understand their roles and significance.

**Independent Variable:**

(1) The independent variable in our study was the 'type of image,' categorized into five levels:
   (a) Camera-generated images
   (b) DALL-E2 generated images
   (c) DALL-E3 generated images
   (d) Stable Diffusion generated images
   (e) GLIDE generated images

**Dependent Variables:**

(1) Photorealism: Assessed using a 5-item questionnaire.

*4.5.1 Statistical Analysis.* In our study, the primary statistical method used was repeated measures Analysis of Variance (ANOVA), which is ideal for comparing means across multiple groups, in this case, different types of images (Camera-generated, DALL-E2, DALL-E3, Stable Diffusion, and GLIDE).

The significance of differences between these groups was further analyzed using Tukey's Honest Significant Difference (HSD) test for post hoc pairwise comparisons. This approach allowed for a detailed examination of how each type of image varied in terms of photorealism, image quality, and text-image alignment, with an emphasis on statistical rigor (significance set at $p < 0.05$) and clarity in the presentation of results (including F values, p-values, mean differences, and confidence intervals).

The analysis of photorealism across different image types using repeated measures ANOVA revealed a highly significant effect ($F = 173.1569$, $p < 0.0001$). This indicates that the type of image significantly influences the perception of photorealism. The mean ratings for photorealism varied notably among the image types, with Camera-generated images generally perceived as more photorealistic.

Table 10. Demographics Respondent Profiles

| ID | Responses | Percentage |
|---|---|---|
| DQ1G | Male | 57% |
| | Female | 42% |
| | Non-binary / Third gender | 1% |
| DQ2G | Under 18 | 2% |
| | 18 - 24 | 38% |
| | 25 - 34 | 38% |
| | 35 - 44 | 14% |
| | 45 - 54 | 5% |
| | 55 - 64 | 2% |
| | 65 - 74 | 1% |
| DQ3G | High School or lower | 27% |
| | Associate's Degree or equivalent | 10% |
| | Bachelor's Degree | 48% |
| | Master's Degree | 13% |
| | Doctorate or higher | 2% |
| DQ4G | I have some experience (have taken courses or dabble in it). | 28% |
| | I'm an expert (work with AI or CGI professionally). | 4% |
| | I've heard of it but don't know much. | 18% |
| | I'm familiar but don't have hands-on experience. | 49% |
| | I'm not familiar at all. | 2% |
| DQ5G | Multiple times a day | 48% |
| | Weekly | 17% |
| | Daily | 30% |
| | Monthly | 4% |
| | Rarely | 2% |
| DQ6G | Yes, as a hobby or pastime. | 28% |
| | No, but I appreciate the arts and often view such content. | 45% |
| | I have taken courses or received training. | 9% |
| | No experience or particular interest. | 12% |
| | Yes, I'm a professional. | 7% |
| DQ7G | Republic of South Africa | 14% |
| | Portuguese Republic | 13% |
| | Republic of Poland | 13% |
| | Islamic Republic of Pakistan | 11% |
| | United Kingdom | 8% |
| | United Mexican States | 6% |
| | Hungary | 3% |
| | Hellenic Republic (Greece) | 4% |
| | Kingdom of Spain | 3% |
| | Republic of Italy | 4% |
| | United States of America | 5% |

*There are less than 2% responses from France, Estonia, Slovenia, France, Czech Republic, Canada, Belgium, Zimbabwe, Germany, India, Portugal, Scotland, etc*

The Tukey's HSD post hoc test provided further insights into pairwise comparisons:

(1) Camera-generated images were consistently rated as more photorealistic compared to all AI-generated images. The greatest difference was observed between Camera Generated and Glide images, with a mean difference of $-2.02$ ($p < 0.001$), indicating a higher rating for Camera Generated images.

(2) Comparisons within AI-generated images also showed significant differences. For instance, DALL-E2 images were rated as more photorealistic compared to Glide images, with a mean difference of $-1.58$ ($p < 0.001$).

Table 11. Participant Responses (Mean (Out of 5))

| Dimension | Camera Gen. | DALL-E2 | GLIDE | Stable Diffusion | DALL-E3 |
|---|---|---|---|---|---|
| Photorealism | 4.06 | 3.63 | 2.04 | 3.30 | 2.75 |
| Image Quality | 3.57 | 3.91 | 2.10 | 3.73 | 3.98 |
| Text-Image Alignment | 4.07 | 3.11 | 2.03 | 3.22 | 3.65 |

Table 12. Response Distribution for Photorealism

| Question ID | Scale | Camera Gen. | DALLE 2 | GLIDE | STABLE DIFUSION | DALLE 3 |
|---|---|---|---|---|---|---|
| PQ1R | Strongly Agree | 49% | 39% | 10% | 35% | 26% |
| | Somewhat Agree | 32% | 25% | 10% | 21% | 13% |
| | Neither agree nor disagree | 5% | 6% | 8% | 6% | 5% |
| | Somewhat disagree | 10% | 19% | 19% | 13% | 12% |
| | Strongly disagree | 4% | 12% | 53% | 25% | 44% |
| PQ2R | Strongly Agree | 56% | 44% | 9% | 36% | 27% |
| | Somewhat Agree | 28% | 26% | 13% | 21% | 16% |
| | Neither agree nor disagree | 7% | 8% | 20% | 8% | 9% |
| | Somewhat disagree | 7% | 13% | 51% | 12% | 11% |
| | Strongly disagree | 2% | 10% | 8% | 23% | 37% |
| PQ3R | Strongly Agree | 45% | 36% | 9% | 34% | 26% |
| | Somewhat Agree | 34% | 24% | 7% | 21% | 15% |
| | Neither agree nor disagree | 9% | 11% | 8% | 8% | 8% |
| | Somewhat disagree | 9% | 17% | 21% | 15% | 12% |
| | Strongly disagree | 2% | 12% | 55% | 21% | 39% |
| PQ4R | Strongly Agree | 41% | 33% | 7% | 31% | 23% |
| | Somewhat Agree | 32% | 25% | 11% | 21% | 13% |
| | Neither agree nor disagree | 12% | 11% | 10% | 11% | 8% |
| | Somewhat disagree | 12% | 15% | 22% | 14% | 13% |
| | Strongly disagree | 4% | 15% | 50% | 23% | 43% |
| PQ5R | Strongly Agree | 38% | 36% | 9% | 33% | 27% |
| | Somewhat Agree | 31% | 29% | 13% | 24% | 18% |
| | Neither agree nor disagree | 14% | 11% | 10% | 9% | 7% |
| | Somewhat disagree | 12% | 14% | 21% | 14% | 10% |
| | Strongly disagree | 5% | 9% | 47% | 20% | 38% |

These results have a significant impact of image type on perceived photorealism, with Camera-generated images consistently rated as more realistic than AI-generated images. Figure 3 illustrates the average ratings with a 95% confidence interval. The observed significant differences among AI-generated images further emphasize the varying capabilities of different generative models in producing photorealistic images.

## 4.6 Statistical Analysis: Image Quality

Interestingly, DALL-E3 led the scores for image quality with a mean of 3.98, slightly surpassing DALL-E2's 3.91. It is important to note that DALL-E 3 was not impressive in photorealism. This suggests that while DALL-E3 may not be the most photorealistic, it excels in producing high-quality images. Stable Diffusion also scored well with 3.73. Which indicates its effectiveness in generating visually appealing images. Camera-generated images, while scoring a solid 3.57, were slightly behind the AI models in this metric, which might suggest that AI models are starting to match or even surpass traditional photography in certain aspects of image quality. It depends upon the quality of the Camera used to capture the image. GLIDE, again, had the lowest score of 2.10 by reinforcing its relative weakness in both realism and quality.

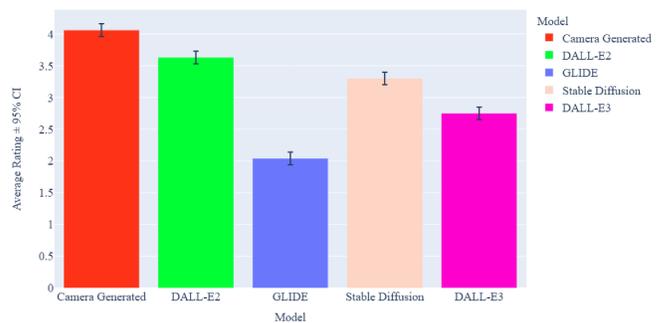

Fig. 3. Photorealsim Ratings: Average Ratings with 95% Confidence Intervals Across Different Model Types

Table 13. Statistical Analysis: Photorealism

| Group 1 | Group 2 | Mean Diff. | p-value | Lower CI | Upper CI | Significance |
|---------|---------|-----------|---------|----------|----------|--------------|
| Camera Generated | DALL-E2 | -0.44 | 0.0010 | -0.70 | -0.17 | Yes |
| Camera Generated | DALL-E3 | 1131 | 0.0010 | -1.58 | -1.04 | Yes |
| Camera Generated | Glide | -2.02 | 0.0010 | -2.29 | -1.75 | Yes |
| Camera Generated | Stable Diffusion | -0.76 | 0.0010 | -1.03 | -0.49 | Yes |
| DALL-E2 | DALL-E3 | -0.88 | 0.0010 | -1.15 | -0.61 | Yes |
| DALL-E2 | Glide | -1.58 | 0.0010 | -1.85 | -1.32 | Yes |
| DALL-E2 | Stable Diffusion | -0.32 | 0.0092 | -0.59 | -0.05 | Yes |
| DALL-E3 | Glide | -0.71 | 0.0010 | -0.98 | -0.44 | Yes |
| DALL-E3 | Stable Diffusion | 0.55 | 0.0010 | 0.28 | 0.82 | Yes |
| Glide | Stable Diffusion | 1.26 | 0.0010 | 0.99 | 1.53 | Yes |

We further analyzed deeper questions, For the first question IQ1G (the image is clear and sharp), 48% of respondents strongly agreed that DALL-E2 images were clear and sharp, followed closely by 46% for Stable Diffusion and 59% for DALL-E3. Camera-generated images had 31% strongly agree, which indicates a high but relatively lower perception compared to some AI models. GLIDE had the lowest, with only 9% strongly agreeing, and 44% strongly disagreeing, suggesting significant issues with image clarity and sharpness. In IQ2G (the colors in the image are vibrant and lifelike), both DALL-E2 and DALL-E3 had 48% of respondents strongly agreeing, indicating superior color vibrancy. Camera-generated images were again rated lower with 27% strongly agreeing. GLIDE's color vibrancy was rated poorly, with only 9% strongly agreeing and 30% strongly disagreeing. For IQ3G (overall satisfaction with the image quality), DALL-E3 led with 46% strongly agreeing, followed by DALL-E2 at 36% and camera-generated images at 30%. GLIDE was the lowest with only 6% strongly agreeing, showing significant dissatisfaction among participants. In IQ5G (resolution meeting expectations), DALL-E3 scored highest again with 42% strongly agreeing, followed by DALL-E2 at 37% and camera-generated images at 29%. GLIDE continued to lag behind, with 54% strongly disagreeing.

These results have varying perceptions of image quality across different types of images. While DALL-E2 and DALL-E3 were often rated higher than camera-generated images in terms of clarity, color vibrancy, and overall quality, GLIDE consistently scored lower which highlights its deficiencies. The data suggests that advancements in AI-generated imagery, particularly with models like DALL-E2 and DALL-E3, are approaching and, in some cases, surpassing the quality perceived in traditional camera-generated images.

Similarly to photorealism analysis, our research question was formulated as follows: Our aim is to identify disparities in image quality among various AI models compared to camera-generated images. Our hypothesis aims to explore the disparities in perceived image quality among different types of images, including those generated by cameras and various AI models such as DALL-E2, DALL-E3, Stable Diffusion, and GLIDE. The objective is to ascertain the extent and nature of differences in photorealism across these distinct image sources.

We also detail the independent and dependent variables used in our study to understand their roles and significance.

**Independent Variable:**

(1) The independent variable in our study was the 'type of image,' categorized into five levels:
   (a) Camera-generated images
   (b) DALL-E2 generated images
   (c) DALL-E3 generated images
   (d) Stable Diffusion generated images
   (e) GLIDE generated images

**Dependent Variables:**

(1) Image Quality: Evaluated through a 4-item questionnaire. (1 item excluded as result of validation)

Similar to photorealism, the ANOVA for image quality showed a significant effect of image type ($F = 180.2068$, $p < 0.0001$). This indicates that the type of image significantly influences the perception of image quality. Camera-generated images were typically perceived as higher in quality than AI-generated images.

The post hoc analysis revealed:

(1) DALL-E3 images had significantly higher image quality ratings compared to all AI-generated types. Notably, the difference between DALL-E3 and Glide was the most pronounced.

Table 14. Response Distribution for Image Quality

| Question ID | Scale | Camera Gen. | DALLE 2 | GLIDE | STABLE DIFUSION | DALLE 3 |
|---|---|---|---|---|---|---|
| IQ1G | Strongly Agree | 31% | 48% | 9% | 46% | 59% |
| | Somewhat Agree | 38% | 32% | 11% | 28% | 26% |
| | Neither agree nor disagree | 10% | 8% | 8% | 8% | 4% |
| | Somewhat disagree | 17% | 8% | 28% | 9% | 6% |
| | Strongly disagree | 3% | 4% | 44% | 10% | 6% |
| IQ2G | Strongly Agree | 27% | 48% | 9% | 46% | 48% |
| | Somewhat Agree | 28% | 30% | 21% | 28% | 25% |
| | Neither agree nor disagree | 13% | 10% | 14% | 8% | 7% |
| | Somewhat disagree | 14% | 8% | 26% | 9% | 11% |
| | Strongly disagree | 19% | 4% | 30% | 10% | 9% |
| IQ3G | Strongly Agree | 30% | 36% | 6% | 36% | 46% |
| | Somewhat Agree | 32% | 28% | 6% | 23% | 26% |
| | Neither agree nor disagree | 19% | 13% | 8% | 11% | 12% |
| | Somewhat disagree | 15% | 15% | 26% | 17% | 10% |
| | Strongly disagree | 5% | 9% | 54% | 13% | 6% |
| IQ5G | Strongly Agree | 29% | 37% | 7% | 34% | 42% |
| | Somewhat Agree | 28% | 28% | 7% | 24% | 22% |
| | Neither agree nor disagree | 18% | 13% | 6% | 14% | 16% |
| | Somewhat disagree | 19% | 15% | 26% | 15% | 10% |
| | Strongly disagree | 6% | 7% | 54% | 13% | 10% |

Table 15. Statistical Analysis: Image Quality

| Group 1 | Group 2 | Mean Diff. | p-value | Lower CI | Upper CI | Significance |
|---|---|---|---|---|---|---|
| Camera Generated | DALL-E2 | -0.96 | 0.0010 | -1.21 | -0.71 | Yes |
| Camera Generated | DALL-E3 | -0.42 | 0.0010 | -0.67 | -0.17 | Yes |
| Camera Generated | Glide | -2.04 | 0.0010 | -2.28 | -1.79 | Yes |
| Camera Generated | Stable Diffusion | -0.85 | 0.0010 | -1.10 | -0.60 | Yes |
| DALL-E2 | DALL-E3 | 0.54 | 0.0010 | 0.29 | 0.79 | Yes |
| DALL-E2 | Glide | -1.08 | 0.0010 | -1.32 | -0.83 | Yes |
| DALL-E2 | Stable Diffusion | 0.11 | 0.7346 | -0.14 | 0.36 | No |
| DALL-E3 | Glide | -1.62 | 0.0010 | -1.87 | -1.37 | Yes |
| DALL-E3 | Stable Diffusion | -0.43 | 0.0010 | -0.68 | -0.19 | Yes |
| Glide | Stable Diffusion | 1.18 | 0.0010 | 0.94 | 1.43 | Yes |

(2) Within AI-generated images, DALL-E2 and DALL-E3 showed similar levels of perceived quality, with no significant difference in their ratings. However, both were rated significantly higher than Glide.

(3) Stable Difusion images are also better than Camera-generated images in terms of quality. This shows that AI-generated commercial models can generate images of very high quality.

These results underscore the considerable influence of image type on perceived photorealism, with DALL-E 3 images consistently rated as more realistic than both Camera-generated and other AI-generated images. Figure 3 visually depicts the average ratings along with a 95% confidence interval. Additionally, the observed significant differences among AI-generated images further highlight the diverse capabilities of different generative models in producing high-quality images.

### 4.7 Statistical Analysis: Text-Image Alignment

Camera-generated images scored highest again in this category with 4.07, reflecting participants' perception that these images best match the provided descriptions. This is expected as these images are inherently aligned with their captions. Among AI models, DALL-E3 performed the best with a score of 3.65, indicating significant advancements in aligning text and images compared to earlier models. DALL-E2 and Stable Diffusion scored 3.11 and 3.22 respectively, showing moderate alignment capabilities. GLIDE's score of 2.03 indicates considerable room for improvement in generating images that accurately reflect textual descriptions.

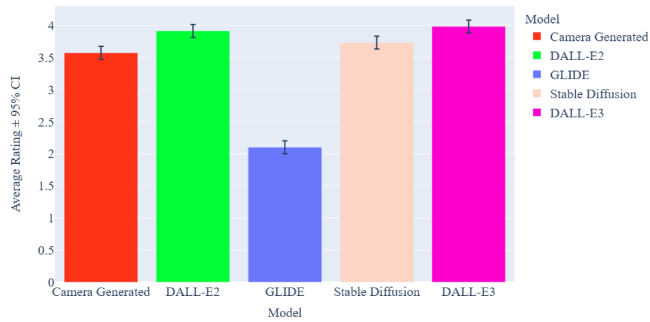

Fig. 4. Image Quality Ratings Across Different Model Types

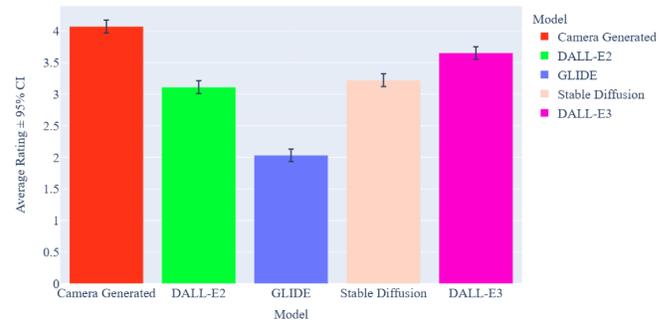

Fig. 5. Text-Image Alignment Ratings Across Different Model Types

For CQ1M (the image perfectly aligns with the given caption), camera-generated images received the highest "Strongly Agree" responses (46%), followed closely by DALL-E3 (42%). GLIDE had the lowest (9%) and the highest "Strongly Disagree" rate (47%), which indicates it was perceived as the least aligned with its captions. Similarly, in CQ2M, camera-generated images were the most aligned contextually (50% "Strongly Agree"), with GLIDE again scoring the lowest (7% "Strongly Agree"). For CQ3M (participants' ability to describe the image with a caption matching the provided one), camera-generated images scored highest (38% "Strongly Agree"), while GLIDE had the highest rate of "Strongly Disagree" responses (47%). This trend continued with CQ4M and CQ5M, where camera-generated images consistently ranked highest, and GLIDE scored the lowest.

We also detail the independent and dependent variables used in our study to understand their roles and significance.

**Independent Variable:**

(1) The independent variable in our study was the 'type of image,' categorized into five levels:
   (a) Camera-generated images
   (b) DALL-E2 generated images
   (c) DALL-E3 generated images
   (d) Stable Diffusion generated images
   (e) GLIDE generated images

**Dependent Variables:**

(1) Text-Image Alignment: Evaluated through a 4-item questionnaire.

The analysis of text-image alignment also indicated a significant effect of image type F Value: 203.94, $p < 0.0001$). This reflects a differential perception of how well the text aligns with the imagery across various image types.

In the post hoc comparisons:

(1) Camera-generated images generally had higher alignment ratings than AI-generated images. The comparison between Camera Generated and Glide images was especially significant, with a mean difference of $-1.46$ ($p < 0.001$).

(2) Among AI-generated images, DALL-E3 was perceived to have slightly better alignment compared to DALL-E2 and significantly better than Glide. However, the difference between DALL-E2 and Stable Diffusion was not significant.

### 4.8 Discussion and Implications

The findings of our study highlight significant differences in human perception of photorealism, image quality, and text-image alignment between camera-generated and AI-generated images.

Camera-generated images were rated as the most photorealistic, with an average score of 4.06 out of 5. They also aligned most closely with their captions, achieving a mean score of 4.07. This result is expected, as camera-generated images are real photographs inherently aligned with their descriptions. DALL-E2 emerged as the second most photorealistic, with a score of 3.63. However, DALL-E3 scored only 2.75, indicating that humans do not consider DALL-E3 images to be as photorealistic as those generated by DALL-E2. The differences in photorealism between DALL-E2 and DALL-E3 can be attributed to their underlying architectures. DALL-E2 utilizes a modified version of GPT-3, a transformer-based architecture, whereas DALL-E3 is based on GPT-4, producing higher-resolution images at 1024x1024 pixels compared to DALL-E2's 512x512 pixels. While the increased resolution allows DALL-E3 to capture more detail and create clearer images, it may also result in the generation of unrealistic scenes. This is particularly evident in question PQ2R, where 37% of participants strongly disagreed that the scenes in DALL-E3 images could exist in the real world.

Moreover, DALL-E3 incorporates more extensive safety measures, potentially influencing the type of content it generates compared to DALL-E2. These safety measures might contribute to a reduced focus on photorealism, even though DALL-E3 excels in generating high-quality images, with an average score of 3.98 [Betker et al. 2023]. Stable Diffusion also received relatively high photorealism scores from participants, averaging 3.30. Its architecture, based on Variational Autoencoders, typically outperforms other generative models in terms of quality. However, achieving realism remains a significant challenge for any generative model due to the potential for generating artifacts based on training data.

GLIDE-generated images were rated lower in photorealism, largely due to their limited generalization capabilities on unseen datasets. Unlike the other models, GLIDE is not commercialized and has been trained on basic datasets, restricting its ability to produce high-quality images beyond its trained capacity. This limitation was evident across all three evaluated categories. Regarding image quality, DALL-E3 performed exceptionally well, with a score of 3.98, reflecting its strong foundation in GPT-4, which supports high-quality image generation through its advanced architecture and extensive training data. However, despite producing high-quality images, DALL-E3 struggles with photorealism. This insight is crucial for the development of commercial models intended for real-world applications.

Table 16. Response Distribution for Text-Image Alignment

| Question ID | Scale | Camera Gen. | DALLE 2 | GLIDE | STABLE DIFFUSION | DALLE 3 |
|---|---|---|---|---|---|---|
| CQ1M | Strongly Agree | 46% | 26% | 9% | 27% | 42% |
| | Somewhat Agree | 31% | 21% | 8% | 28% | 28% |
| | Neither agree nor disagree | 9% | 6% | 8% | 11% | 9% |
| | Somewhat disagree | 10% | 28% | 28% | 17% | 11% |
| | Strongly disagree | 3% | 20% | 47% | 17% | 10% |
| CQ2M | Strongly Agree | 50% | 25% | 7% | 28% | 41% |
| | Somewhat Agree | 34% | 27% | 12% | 28% | 28% |
| | Neither agree nor disagree | 6% | 10% | 9% | 9% | 10% |
| | Somewhat disagree | 8% | 25% | 32% | 18% | 11% |
| | Strongly disagree | 2% | 13% | 40% | 17% | 9% |
| CQ3M | Strongly Agree | 38% | 22% | 9% | 24% | 32% |
| | Somewhat Agree | 37% | 25% | 7% | 27% | 27% |
| | Neither agree nor disagree | 12% | 13% | 10% | 11% | 11% |
| | Somewhat disagree | 10% | 21% | 27% | 20% | 16% |
| | Strongly disagree | 3% | 18% | 47% | 18% | 13% |
| CQ5M | Strongly Agree | 40% | 21% | 7% | 21% | 32% |
| | Somewhat Agree | 35% | 22% | 8% | 23% | 24% |
| | Neither agree nor disagree | 12% | 12% | 8% | 15% | 16% |
| | Somewhat disagree | 10% | 28% | 28% | 22% | 16% |
| | Strongly disagree | 3% | 17% | 50% | 20% | 12% |

Table 17. Text-Image Alignment

| Group 1 | Group 2 | Mean Diff. | p-value | Lower CI | Upper CI | Significant |
|---|---|---|---|---|---|---|
| Camera Gen. | DALL-E2 | 0.34 | 0.0010 | 0.11 | 0.57 | Yes |
| Camera Generated | DALL-E3 | 0.41 | 0.0010 | 0.18 | 0.64 | Yes |
| Camera Generated | Glide | -1.46 | 0.0010 | -1.69 | -1.24 | Yes |
| Camera Generated | Stable Diffusion | 0.16 | 0.3153 | -0.07 | 0.39 | No |
| DALL-E2 | DALL-E3 | 0.07 | 0.9000 | -0.16 | 0.30 | No |
| DALL-E2 | Glide | -1.80 | 0.0010 | -2.03 | -1.57 | Yes |
| DALL-E2 | Stable Diffusion | -0.18 | 0.2052 | -0.41 | 0.05 | No |
| DALL-E3 | Glide | -1.87 | 0.0010 | -2.10 | -1.65 | Yes |
| DALL-E3 | Stable Diffusion | -0.25 | 0.0243 | -0.48 | -0.02 | Yes |
| Glide | Stable Diffusion | 1.62 | 0.0010 | 1.40 | 1.85 | Yes |

Stable Diffusion also garnered high trust from participants, achieving a score of 3.73 for image quality. This performance is attributed to its advanced architecture and internal structure. Camera-generated images ranked third in this category, with a score of 3.57, followed by GLIDE, which again failed to meet human expectations. The trend indicates that AI-generated images often show higher quality than original camera-generated images. This trend is likely due to the success of neural networks, including super-resolution GANs and other AI models, in effectively denoising low-quality images and producing high-quality visuals. The capabilities of DALL-E2,

DALL-E3, and Stable Diffusion in generating high-quality images have gained significant trust from participants.

In terms of text-image alignment, camera-generated images naturally scored the highest, with an average of 4.07, due to their inherent alignment with real-world scenes. Among the AI models, DALL-E3 showed notable improvements over its predecessors, scoring 3.65, an increase from DALL-E2's score of 3.11. This improvement suggests advancements in DALL-E3's ability to accurately interpret and visualize textual descriptions. The enhanced capabilities of GPT-4 contribute to DALL-E3's improved performance in generating well-aligned images. In contrast, GLIDE once again lagged behind, highlighting its challenges in accurately translating text descriptions into corresponding visual representations.

It is also noteworthy that the statistical properties of camera-generated images, such as hue, saturation, brightness, and vibrancy, are generally lower than those of AI-generated images. However, entropy levels are nearly similar for both AI and camera-generated images. This similarity in entropy is a promising indication for designing effective image metrics. The differences in hue, saturation, brightness, and vibrancy suggest that AI-generated images have distinct characteristics that can be quantitatively measured and used to develop metrics for distinguishing between AI and camera-generated images.

## 5 STUDY 3: BENCHMARKING COMPUTATIONAL IMAGE METRICS FOR PHOTOREALISM, IMAGE QUALITY, AND TEXT-IMAGE ALIGNMENT AGAINST HUMAN PERCEPTION

This section details the assessment of traditional metrics for key dimensions such as Photorealism, Image Quality, and Text-Image Alignment. Additionally, it includes the design and proposal of the Neural Feature Similarity Score (NFSS) for image quality. An expert evaluation of the Interpolative Binning Scale (IBS), a scaling method used to compare human and metric scores, is also presented.

### 5.1 Interpolative Binning Scale (IBS)

To bridge the gap between human evaluative judgments and numerical metric scores, we propose the Interpolative Binning Scale (IBS) [Aziz et al. 2024], an innovative quantitative scaling approach. The IBS aims to translate metric scores into a format that is both interpretable and reflective of human assessment patterns. This is achieved by initially grouping metric scores into predefined categories that denote interpretative meaning, followed by refining these groupings through interpolation to ascertain the precise position of each score within its category. This dual-phase methodology facilitates a granular transformation of raw metric values into a scoring framework akin to human evaluative gradations.

The scores of various metrics were presented to the experts of the Human-Centric Computing Group at Western University [Human Centric Computing Group 2024]. These experts categorized the scores into bins based on their interpretative meanings, thereby defining the Interpolative Binning Scale (IBS). The classification involved assigning metric scores to discrete categories that ranged from 'Strongly Disagree' to 'Strongly Agree', akin to the Likert scale used in human evaluations. This approach ensured that the metric scores were aligned with human perceptions of image quality, enhancing the interpretability and relevance of the evaluation process. By employing the IBS, each metric provides a robust and nuanced evaluation of image quality that closely mirrors human judgment.

*5.1.1 Classification and Interpolation.* The initial phase of the IBS involves the classification of raw metric scores into distinct bins representing levels of agreement or quality, such as 'Strongly Agree' to 'Strongly Disagree'. This transformation converts the continuous range of metric scores into a set of discrete, interpretable categories. The rationale behind this approach is that human evaluative scores often utilize a Likert scale, ranging from 'Strongly Agree' to 'Strongly Disagree'. By mapping metric scores into these bins, we create a more relatable and understandable framework.

In the subsequent interpolation phase, each classified score is assigned an exact numerical value or position within its bin. This process employs linear interpolation, allowing for the determination of the precise score. Specifically, for a given metric score $x$, and a sequence of ordered bins $\{b_i\}$ with associated scores $\{s_i\}$, the IBS score is calculated as follows:

$$\text{IBS Score} = s_i + \left( \frac{s_{i+1} - s_i}{b_{i+1} - b_i} \right) \times (x - b_i) \qquad (10)$$

In this equation, $s_{i+1}$ and $s_i$ are the scores corresponding to the bins immediately above and below the value $x$, respectively, while $b_{i+1}$ and $b_i$ represent the boundaries of the bin within which $x$ falls. This interpolation method provides a robust and interpretable means to compare human and metric scores effectively.

*5.1.2 Illustrative Example.* In the context of image quality evaluation, let's consider the Structural Similarity Index, a crucial metric that necessitates qualitative interpretation. For this illustrative example, we will scale an SSIM score of 0.45 using the proposed Interpolative Binning Scale method.

*Classification.* Initially, the SSIM score is classified into one of the predefined categories that represent various levels of image quality. These categories, or bins, are detailed in Table 18. For an SSIM score of 0.45, the classification falls into the 'Somewhat Agree' bin.

*Interpolation.* The next step involves interpolating the score within this bin. The formula used for interpolation within the 'Somewhat Agree' range is as follows:

$$\text{IBS Score} = 3.1 + \left( \frac{4.1 - 3.1}{0.6 - 0.3} \right) \times (0.45 - 0.3) \qquad (11)$$

By calculating the above formula, we derive the interpolated IBS score:

$$\text{IBS Score} = 3.1 + \left( \frac{1}{0.3} \right) \times (0.15) = 3.1 + 0.45 = 3.55 \qquad (12)$$

*Interpretation.* The IBS score of 3.55 indicates that the original SSIM value of 0.45 aligns closer to the 'Somewhat Agree' end of the score spectrum. This score provides a refined and precise placement of the SSIM value within its bin, enhancing interpretability.

This scoring method offers significant advantages over traditional normalization techniques, such as Z-Standardization and Min-Max scaling, as well as linear regression-based black-box methods. It mitigates inherent biases and provides a more interpretable score that is comprehensible to humans, thereby facilitating a better comparison between human and metric evaluations.

Table 18. Classification bins for Metrics

| Criteria | Score | SSIM | PSNR | FID | NFSS | LPIPS | IS |
|---|---|---|---|---|---|---|---|
| Strongly Disagree | 0.0 - 1.0 | -1 to -0.6 | 0-7 | > 150 | -1 to -0.6 | 0.9 to 1 | 0 < 1 |
| Somewhat Disagree | 1.1 - 2.0 | -0.5 to -0.2 | 8-15 | 100 to 149 | -0.5 to -0.2 | 0.7 to 0.8 | 1 < 2 |
| Neutral | 2.1 - 3.0 | -0.1 to 0.2 | 16-23 | 31 to 99 | -0.1 to 0.2 | 0.5 to 0.6 | 2 < 3 |
| Somewhat Agree | 3.1 - 4.0 | 0.3 to 0.6 | 24-31 | 11 to 30 | 0.3 to 0.6 | 0.3 to 0.4 | 3 < 5 |
| Strongly Agree | 4.1 - 5.0 | 0.7 to 1.00 | > 32 | < 10 | 0.7 to 1.00 | 0 to 0.2 | > 6 |

## 5.2 Neural Features Similarity Score (NFSS)

The rapid advancement in image generation models necessitates robust and comprehensive metrics for evaluating image quality. The Neural Features Similarity Score (NFSS) was designed to address these needs by integrating structural similarity, perceptual similarity, and color histogram similarity into a single, unified metric. This multi-dimensional approach ensures that NFSS can robustly evaluate image quality in a manner that is both scientifically rigorous and perceptually relevant. The first component of NFSS is the Multi-Scale Structural Similarity Index (MS-SSIM). MS-SSIM measures the structural similarity between two images across multiple scales, providing a comprehensive assessment of structural integrity. It divides images into patches of varying sizes and computing the SSIM for each patch, MS-SSIM captures both fine and coarse structural details. The formula for multi-scale SSIM is given by:

$$\text{SSIM}_{\text{MS}} = \frac{1}{n} \sum_{i=1}^{n} \text{SSIM}(P_i, Q_i) \qquad (13)$$

where $P_i$ and $Q_i$ are patches of the images being compared, and $n$ is the number of patches.

The second component of NFSS is the Perceptual Distance Metric (PDM). PDM evaluates the perceptual similarity between two images using deep neural network features. This method captures high-level perceptual features that are important for human visual perception but may not be captured by traditional pixel-based metrics. The formula for perceptual distance is:

$$\text{D}_{\text{P}}(I_1, I_2) = \|\phi(I_1) - \phi(I_2)\| \qquad (14)$$

where $\phi$ denotes the feature extraction function of the pre-trained network, and $I_1$ and $I_2$ are the images being compared.

The third component of NFSS is the Color Histogram Similarity (CHS). CHS assesses the similarity based on the color distribution of the images. This method captures color information that is crucial for evaluating the overall appearance and quality of images. The formula for color histogram similarity is:

$$\text{H}_{\text{C}}(I_1, I_2) = \frac{\sum_{i=1}^{k} (H_{I_1,i} - \bar{H}_{I_1})(H_{I_2,i} - \bar{H}_{I_2})}{\sqrt{\sum_{i=1}^{k} (H_{I_1,i} - \bar{H}_{I_1})^2 \sum_{i=1}^{k} (H_{I_2,i} - \bar{H}_{I_2})^2}} \qquad (15)$$

where $H_{I_1,i}$ and $H_{I_2,i}$ are the histogram values of the images $I_1$ and $I_2$ for bin $i$, and $\bar{H}_{I_1}$ and $\bar{H}_{I_2}$ are the mean histogram values.

The entropy component measures the randomness or complexity of the image data. Higher entropy usually indicates more complex texture or content in the image. The entropy formula is:

$$H(I) = -\sum_{i=0}^{255} p_i \log_2(p_i + \epsilon) \qquad (16)$$

where $p_i$ is the normalized histogram of the grayscale image $I$, and $\epsilon$ is a small value to avoid $\log(0)$.

The dynamic weighting factor $\alpha$ adjusts the influence of structural versus perceptual similarities based on entropy and color similarity. This ensures that the metric adapts to the characteristics of the images being compared. The formula for alpha is:

$$\alpha = \frac{1}{1 + (\exp(-H + (\text{SSIM}_{\text{MS}} - \text{D}_{\text{P}})) \cdot \exp(-\text{H}_{\text{C}}))} \qquad (17)$$

The final NFSS is computed as a weighted sum of the MS-SSIM score, PDM score, and a scaled colour histogram similarity:

$$\text{NFSS} = (\alpha \cdot \text{SSIM}_{\text{MS}}) + ((1 - \alpha) \cdot \text{D}_{\text{P}}) + (0.1 \cdot \text{H}_{\text{C}}) \qquad (18)$$

By combining these different aspects, NFSS provides a robust and comprehensive evaluation of image quality, capturing both detailed structural information and high-level perceptual features. This metric is designed to align closely with human judgments of image quality, making it a valuable tool for assessing AI-generated images.

## 5.3 Computational Image Metrics Assessment

In this subsection, we will evaluate computational image metrics used to assess realism, image quality, and text-image alignment in images. This evaluation aims to understand how closely these computational image metrics align with human assessments, as human perception is ultimately the gold standard for determining whether an image appears real or fake, its quality, and its suitability in text-image alignment. The selected metrics span a range of technologies and methods to ensure comprehensive coverage. For photorealism, we use Fréchet Inception Distance (FID), Kernel Inception Distance (KID), Learned Perceptual Image Patch Similarity (LPIPS), Structural

Similarity Index (SSIM), and Multi-Scale Structural Similarity Index (MS-SSIM). For image quality, we include FID, Peak Signal-to-Noise Ratio (PSNR), KID, LPIPS, SSIM, FID, Inception Score (IS), and MS-SSIM. We also proposed a hybrid metric namely Neural Feature Similarity Score (NFSS). For text-image alignment, metrics such as CLIP and BERT are employed.

Both camera-generated and AI-generated images, previously discussed in Studies 1 and 2, will be used for this evaluation. These images were initially presented to humans for assessment, and now they will be passed through the computational metrics to assess photorealism, image quality, and text-image alignment. The scores obtained from these metrics will then be rescaled using the Interpolative Binning Scale (IBS) method described in Section 5.1.1. This rescaling will map the metric scores to a 0 to 5 range, ensuring comparability with human scores, which are already in this range as given in Table 11.

To compare the human and metric outputs after IBS scaling, we will use two metrics: Mean Absolute Difference (MAD) and Mean Absolute Percentage Error (MAPE). These metrics provide a clear and interpretable measure of the differences between human and computational metric outputs. Since the comparison is direct, with both sets of scores scaled between 0 and 5, correlation analysis such as Pearson correlation is unnecessary. Instead, we focus on direct analysis by providing absolute numerical differences and percentages, which are easily understood. The formulas for MAD and MAPE are provided below:

$$ \text{MAD} = \frac{1}{n} \sum_{i=1}^{n} |x_i - y_i| \qquad (19) $$

Where $x_i$ represents the human score for the $i$-th image, $y_i$ represents the metric score for the same image, and $n$ is the total number of images.

$$ \text{MAPE} = \frac{100\%}{n} \sum_{i=1}^{n} \left| \frac{x_i - y_i}{x_i} \right| \qquad (20) $$

Here, $x_i$ and $y_i$ retain their meanings as in the MAD equation, providing a percentage error that quantifies the average deviation of metric scores from human scores as a percentage of the human scores.

By applying these metrics, we aim to rigorously quantify the alignment between human and computational assessments of image realism, quality, and text-image alignment by providing valuable insights into the effectiveness and reliability of computational image metrics.

### 5.3.1 Analysis of Photorealism Metrics.
The comparison between metric scores and human judgments on photorealism is detailed in Table 19 and Figure ??. The "Raw Score" column shows the direct output from the metrics, while the "Scaled Score" column represents these scores after being adjusted using the Interpolative Binning Scale (IBS) technique. The categories means likert for both metric and human scores offer a direct and understandable subjective comparison. Additionally, the Mean Absolute Difference (MAD) quantifies the absolute difference between the human and metric scores, and the Mean Absolute Percentage Error (MAPE) expresses the error as a percentage.

For the FID metric, there is a significant deviation from human judgment across various models. The MAPE for FID stands at 37.0% for Stable Diffusion, 35.0% for DALLE2, sharply increases to 130.8% for GLIDE, and remains at 52.7% for DALLE3. These figures indicate that FID scores deviate substantially—by at least 30%—from human scores. Moreover, a comparative analysis of Likert categories reveals that FID consistently fails to align with human-assessed categories. While humans rated images from Stable Diffusion and DALLE2 as 'Somewhat Agree', GLIDE and DALLE3 were rated as 'Neutral'. This inconsistency highlights FID's limitations in capturing the essence of image quality as perceived by humans.

The LPIPS metric shows a better alignment with human judgment compared to FID. Specifically, LPIPS achieves a MAPE of 37.9% for Stable Diffusion, improving over FID's performance. For DALLE2, the error reduces to 26.4%, and for GLIDE, it decreases to 29.9%, which is nearly a 100% improvement compared to FID. Most notably, LPIPS records a minimal error of 16.7% for DALLE3, the lowest among all metrics evaluated for this model. This metric matches the human judgment of 'Neutral' for both GLIDE and DALLE3, demonstrating its superior capability in capturing nuances more reflective of human perception than FID.

SSIM shows a promising correlation with human perception. For instance, the SSIM for Stable Diffusion showed a rescaled score of 3.9, closely aligning with the human score of 3.3, reflecting a high degree of perceptual alignment as both fall into the 'Somewhat Agree' category on the Likert scale with just 17.6% error. However, notable discrepancies emerge in cases such as GLIDE, where despite a high SSIM rescaled score of 4.5 indicating 'Strongly Agree', the corresponding human score was only 2.0 'Neutral', resulting in a substantial MAPE of 120.1%. This significant variance in MAPE highlights the occasional challenges SSIM faces in mirroring human judgments across different settings, suggesting areas for further refinement.

MS-SSIM offers an extended analysis of image quality by assessing visual details across multiple scales. As shown in Table 19, the alignment between the MS-SSIM scores and human judgments varies, but it shows impressive improvement over SSIM alone. For instance, in the case of Stable Diffusion, the MS-SSIM score closely mirrors the human score (3.8 vs. 3.3), and both are categorized as 'Somewhat Agree', indicating a consistent perception of image quality. This scenario exhibits a relatively low MAPE of 15.8%, suggesting a good match between the metric and human perception. Similarly, there is a low error for DALLE2 of just 5.5%. However, GLIDE shows a divergence; despite an MS-SSIM score indicating 'Somewhat Agree' (3.8), the human judgment falls under 'Neutral' (2.0), resulting in a significantly higher MAPE of 88.2%.

We implemented KID with three different kernels, namely Polynomial, RBF, and Exponential. The KID metric shows varied accuracy when compared to human judgments. For instance, in the case of Stable Diffusion, a misalignment is evident with a high MAPE of 69.7%, where the metric categorizes image quality as 'Strongly Disagree', starkly contrasting the human judgment categorized under 'Somewhat Agree'. Similarly, for DALLE2, the discrepancy remains significant with a MAPE of 68.3%, again reflecting a serious mismatch in perceptual evaluation as the metric suggests 'Somewhat Disagree' versus the human-assigned 'Somewhat Agree'. The model GLIDE shows a MAPE of 50.5%, where KID slightly improves in capturing human perception, aligning 'Somewhat Agree' with a 'Neutral' human rating. The smallest MAPE observed is with DALLE3 at 57.8%, where both metric and human judgment categorize the image quality under a neutral standpoint. These results underscore the varying degrees of alignment between computational image metrics and human perception of photorealism. While some metrics like LPIPS and MS-SSIM show closer alignment with human judgments, others like FID and KID shows significant deviations. This analysis highlights the need for continued refinement of these metrics to better capture the subtleties of human visual perception.

Table 19. Comparison of Metric Scores with Human Judgment on Photorealism

| Model | Metric | Raw score | Scaled score | Human score | Category Metric | Category Human | MAD | MAPE |
|-------|--------|-----------|--------------|-------------|-----------------|----------------|-----|------|
| Stable Diffusion | FID | 28.83 | 4.52 | 3.30 | Strongly Agree | Somewhat Agree | 1.22 | 36.96 |
| DALLE2 | | 13.81 | 4.90 | 3.63 | Strongly Agree | Somewhat Agree | 1.27 | 34.98 |
| GLIDE | | 21.28 | 4.71 | 2.04 | Strongly Agree | Neutral | 2.67 | 130.80 |
| DALLE3 | | 41.81 | 4.20 | 2.75 | Strongly Agree | Neutral | 1.45 | 52.72 |
| Stable Diffusion | LPIPS | 0.78 | 2.05 | 3.30 | Neutral | Somewhat Agree | 1.25 | 37.87 |
| DALLE2 | | 0.66 | 2.67 | 3.63 | Neutral | Somewhat Agree | 0.96 | 26.44 |
| GLIDE | | 0.67 | 2.65 | 2.04 | Neutral | Neutral | 0.60 | 29.90 |
| DALLE3 | | 0.74 | 2.29 | 2.75 | Neutral | Neutral | 0.45 | 16.72 |
| Stable Diffusion | SSIM | 0.153 | 3.88 | 3.30 | Somewhat Agree | Somewhat Agree | 0.58 | 17.57 |
| DALLE2 | | 0.195 | 3.98 | 3.63 | Somewhat Agree | Somewhat Agree | 0.35 | 9.64 |
| GLIDE | | 0.396 | 4.49 | 2.04 | Strongly Agree | Neutral | 2.45 | 120.09 |
| DALLE3 | | 0.115 | 3.78 | 2.75 | Somewhat Agree | Neutral | 1.02 | 37.45 |
| Stable Diffusion | MS-SSIM | 0.128 | 3.82 | 3.30 | Somewhat Agree | Somewhat Agree | 0.52 | 15.75 |
| DALLE2 | | 0.134 | 3.83 | 3.63 | Somewhat Agree | Somewhat Agree | 0.20 | 5.50 |
| GLIDE | | 0.139 | 3.84 | 2.04 | Somewhat Agree | Neutral | 1.79 | 88.23 |
| DALLE3 | | 0.058 | 3.64 | 2.75 | Somewhat Agree | Neutral | 0.89 | 32.36 |
| Stable Diffusion | KID | 0.98 | 1 | 3.30 | Strongly Disagree | Somewhat Agree | 2.30 | 69.69 |
| DALLE2 | | 0.67 | 1.15 | 3.63 | Somewhat Disagree | Somewhat Agree | 2.48 | 68.31 |
| GLIDE | | 0.33 | 3.07 | 2.04 | Somewhat Agree | Neutral | 1.02 | 50.49 |
| DALLE3 | | 0.66 | 1.16 | 2.75 | Somewhat Disagree | Neutral | 1.59 | 57.81 |

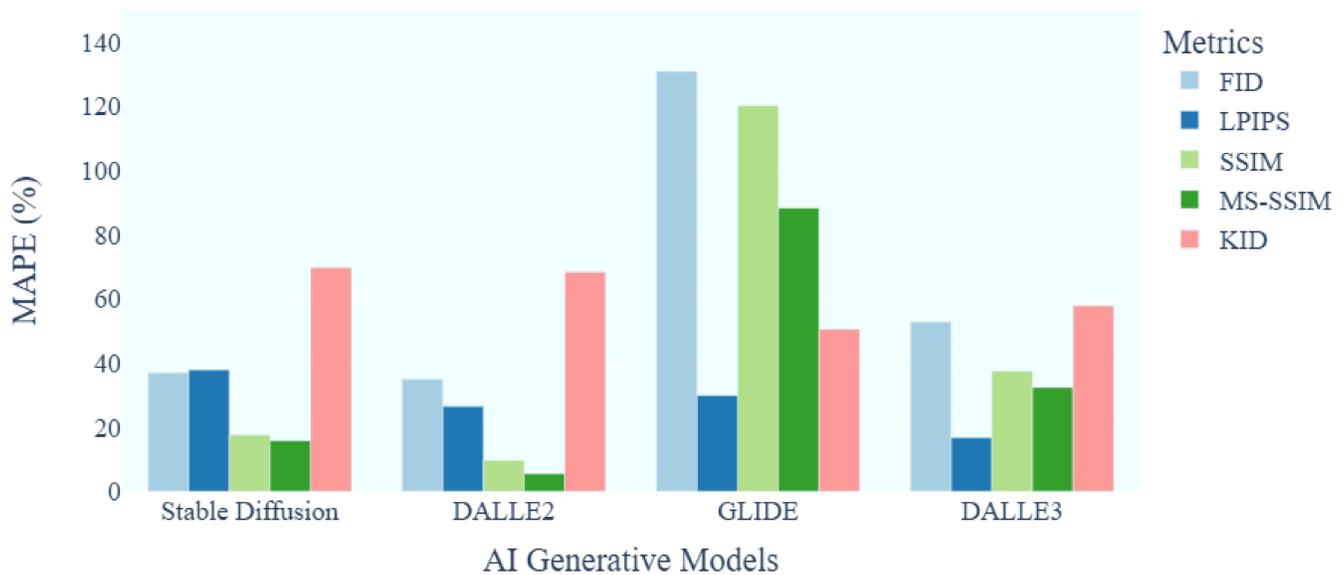

Fig. 6. MAPE Scores of Photorealsim Metrics Across AI Generative Models

*5.3.2 Analysis of Image Quality Metrics.* The comparison between metric scores and human judgments on image quality is detailed in Table 20, Table 21, and Figure ??. The FID metric showed substantial deviations from human judgment, particularly in the case of GLIDE generated image, where the Mean Absolute Percentage Error reached a staggering 124.28%. This error indicates that FID struggles to align with human perception for this model. For Stable Diffusion and DALLE2, the MAPE was 21.17% and 25.31% respectively, suggesting a moderate level of agreement. Interestingly, DALLE3 had the lowest MAPE of 5.52%, indicating a closer alignment with human judgment. Overall, while FID is a useful metric, its reliability varies across different AI generated models.

LPIPS demonstrated a better correlation with human judgments compared to FID. For instance, the MAPE for Stable Diffusion and DALLE2 was 45.04% and 31.71% respectively, indicating a relatively closer alignment. The lowest MAPE of 20.75% was observed for GLIDE, showing LPIPS's ability to better reflect human perception for this model. However, the MAPE for DALLE3 was 42.46%, which is relatively high, suggesting that while LPIPS improves upon FID, it still faces challenges in consistently aligning with human judgment across all models. SSIM showed promising results in correlating with human perception. The MAPE for Stable Diffusion was 4.02%, indicating a high degree of alignment as both scores fall into the 'Somewhat Agree' category. DALLE2 also showed a low MAPE of 1.79%, reflecting a strong alignment with human scores. However, significant discrepancies were observed with GLIDE, where the MAPE was 113.80%, it is highlighting a considerable mismatch. This indicates that while SSIM generally performs well, it can sometimes fail to capture the nuances of human perception.

MS-SSIM extended the analysis of image quality across multiple scales and showed a consistent alignment with human judgments for certain models. For example, Stable Diffusion had a low MAPE of

2.41%, indicating a close match. DALLE2 also showed a low MAPE of 2.05%, reflecting good alignment. However, GLIDE's MAPE was 83.19%, indicating significant divergence. This metric performed well overall, but like SSIM, it can occasionally fail to reflect human perception accurately, particularly for more challenging models. The KID metric showed varied accuracy when compared to human judgments. For Stable Diffusion, the MAPE was 73.19%, indicating a substantial misalignment, as the metric rated image quality much lower than humans. DALLE2 had a MAPE of 70.58%, reflecting a similar mismatch. The MAPE for GLIDE was 46.19%, showing some improvement but still indicating significant discrepancies. DALLE3 had a MAPE of 70.85%, again suggesting a poor alignment with human perception. These results indicate that KID may not be as reliable in capturing human judgments of image quality.

VIF also demonstrated substantial discrepancies from human judgments across all models. For instance, the MAPE for Stable Diffusion was 73.19%, indicating a significant misalignment. DALLE2 showed a similar trend with a MAPE of 74.42%. GLIDE had a MAPE of 52.38%, which, while better, still indicates a considerable gap. DALLE3's MAPE was 74.87%, again reflecting poor alignment. These high errors suggest that VIF may not be an optimal metric for assessing image quality from a human perspective. The Inception Score (IS) showed varying degrees of alignment with human judgments. Stable Diffusion had a MAPE of 44.50%, indicating a moderate level of agreement. DALLE2 performed better with a MAPE of 26.59%, suggesting a closer alignment. However, GLIDE showed a high MAPE of 90.47%, reflecting a substantial misalignment. DALLE3 had a MAPE of 27.63%, indicating a relatively good match with human judgments. IS shows potential but also highlights the need for complementary metrics to fully capture human perceptions of image quality. The NFSS demonstrated a much closer alignment with human judgments across all models. For Stable Diffusion, the

Table 20. Comparison of Metric Scores with Human Judgment on Image Quality (Part 1)

| Model | Metric | Raw score | Scaled score | Human score | Category Metric | Category Human | MAD | MAPE |
|---|---|---|---|---|---|---|---|---|
| Stable Diffusion | FID | 28.83 | 4.52 | 3.73 | Strongly Agree | Somewhat Agree | 0.78 | 21.17 |
| DALLE2 | | 13.81 | 4.90 | 3.91 | Strongly Agree | Somewhat Agree | 0.99 | 25.31 |
| GLIDE | | 21.28 | 4.71 | 2.10 | Strongly Agree | Neutral | 2.61 | 124.28 |
| DALLE3 | | 41.81 | 4.20 | 3.98 | Strongly Agree | Somewhat Agree | 0.22 | 5.52 |
| Stable Diffusion | LPIPS | 0.78 | 2.05 | 3.73 | Neutral | Somewhat Agree | 1.68 | 45.04 |
| DALLE2 | | 0.66 | 2.67 | 3.91 | Neutral | Somewhat Agree | 1.24 | 31.71 |
| GLIDE | | 0.67 | 2.65 | 2.10 | Neutral | Neutral | 0.54 | 20.75 |
| DALLE3 | | 0.74 | 2.29 | 3.98 | Neutral | Somewhat Agree | 1.69 | 42.46 |
| Stable Diffusion | SSIM | 0.15 | 3.88 | 3.73 | Somewhat Agree | Somewhat Agree | 0.14 | 4.02 |
| DALLE2 | | 0.20 | 3.98 | 3.91 | Somewhat Agree | Somewhat Agree | 0.06 | 1.79 |
| GLIDE | | 0.40 | 4.49 | 2.10 | Strongly Agree | Neutral | 2.39 | 113.80 |
| DALLE3 | | 0.12 | 3.78 | 3.98 | Somewhat Agree | Somewhat Agree | 0.20 | 5.02 |
| Stable Diffusion | MS-SSIM | 0.13 | 3.82 | 3.73 | Somewhat Agree | Somewhat Agree | 0.09 | 2.41 |
| DALLE2 | | 0.13 | 3.83 | 3.91 | Somewhat Agree | Somewhat Agree | 0.08 | 2.05 |
| GLIDE | | 0.14 | 3.84 | 2.10 | Somewhat Agree | Neutral | 1.74 | 83.19 |
| DALLE3 | | 0.06 | 3.64 | 3.98 | Somewhat Agree | Somewhat Agree | 0.34 | 8.54 |
| Stable Diffusion | KID | 0.98 | 1.00 | 3.73 | Strongly Disagree | Somewhat Agree | 2.73 | 73.19 |
| DALLE2 | | 0.67 | 1.15 | 3.91 | Somewhat Disagree | Somewhat Agree | 2.76 | 70.58 |
| GLIDE | | 0.33 | 3.07 | 2.10 | Somewhat Agree | Neutral | 0.96 | 46.19 |
| DALLE3 | | 0.66 | 1.16 | 3.98 | Somewhat Disagree | Somewhat Agree | 2.82 | 70.85 |

NFSS had a MAPE of 11.26%, indicating a high degree of agreement. DALLE2 achieved an exceptionally low MAPE of 0.25%, reflecting a nearly perfect match. GLIDE showed a MAPE of 42.85%, which, while higher than the other models, still represents an improvement over VIF and IS. DALLE3 had a MAPE of 11.55%, again indicating a strong alignment with human judgments. These results suggest that NFSS provides a more accurate and reliable measure of image quality from a human perspective compared to VIF and IS.

Table 21. Comparison of Metric Scores with Human Judgment on Image Quality (Part 2)

| Model | Metric | Raw score | Scaled score | Human score | Category Metric | Category Human | MAD | MAPE |
|---|---|---|---|---|---|---|---|---|
| Stable Diffusion | VIF | 0.0041 | 1.00 | 3.73 | Strongly Disagree | Somewhat Agree | 2.73 | 73.19 |
| DALLE2 | | 0.0046 | 1.00 | 3.91 | Strongly Disagree | Somewhat Agree | 2.91 | 74.42 |
| GLIDE | | 0.0058 | 1.00 | 2.10 | Strongly Disagree | Neutral | 1.10 | 52.38 |
| DALLE3 | | 0.0016 | 1.00 | 3.98 | Strongly Disagree | Somewhat Agree | 2.98 | 74.87 |
| Stable Diffusion | IS | 1.04 | 2.04 | 3.73 | Somewhat Disagree | Somewhat Agree | 1.66 | 44.50 |
| DALLE2 | | 1.87 | 2.87 | 3.91 | Somewhat Disagree | Somewhat Agree | 1.04 | 26.59 |
| GLIDE | | 3.00 | 4.00 | 2.10 | Somewhat Agree | Neutral | 1.90 | 90.47 |
| DALLE3 | | 1.88 | 2.88 | 3.98 | Neutral | Somewhat Agree | 1.10 | 27.63 |
| Stable Diffusion | NFSS | 0.17 | 4.15 | 3.73 | Strongly Disagree | Somewhat Agree | 0.42 | 11.26 |
| DALLE2 | | 0.22 | 3.90 | 3.91 | Somewhat Agree | Somewhat Agree | 0.01 | 0.25 |
| GLIDE | | 0.39 | 3.00 | 2.10 | Neutral | Neutral | 0.89 | 42.85 |
| DALLE3 | | 0.11 | 4.44 | 3.98 | Strongly Agree | Somewhat Agree | 0.46 | 11.55 |

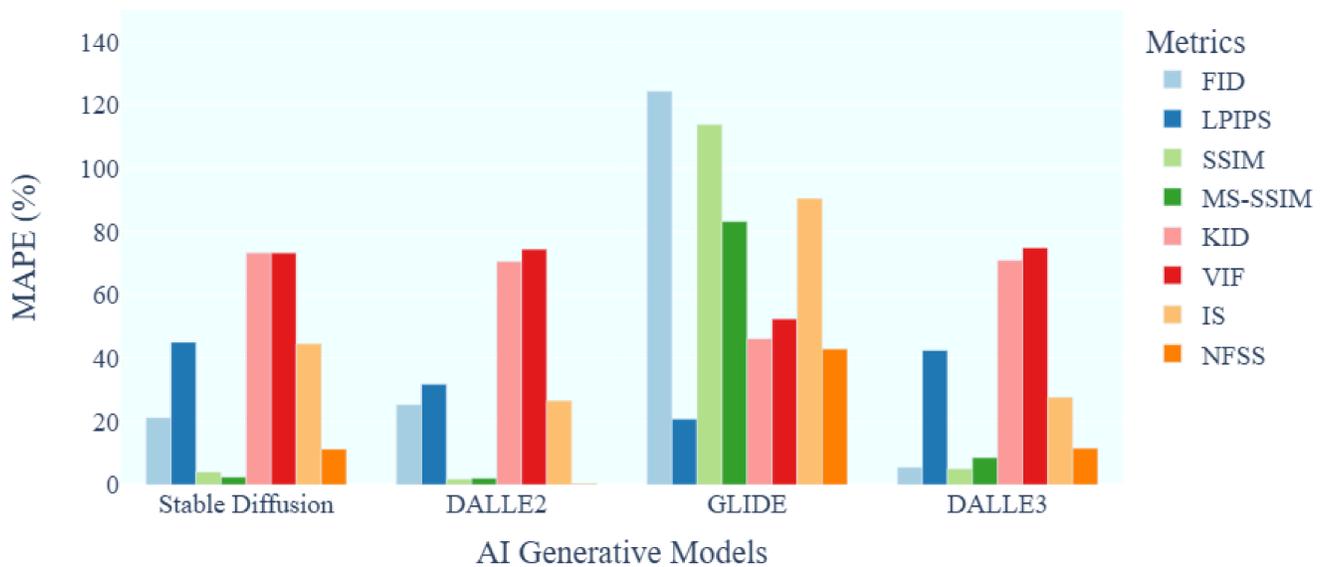

Fig. 7. MAPE Scores of Image Quality Metrics Across AI Generative Models

Table 22. Comparison of Metric Scores with Human Judgment on Text-Image Alignment

| Model | Metric | Raw score | Scaled score | Human score | Category Metric | Category Human | MAD | MAPE |
|---|---|---|---|---|---|---|---|---|
| Stable Diffusion | CLIP | 0.31 | 3.28 | 3.22 | Somewhat Agree | Somewhat Agree | 0.05 | 1.86 |
| DALLE2 | | 0.30 | 3.25 | 3.11 | Somewhat Agree | Somewhat Agree | 0.14 | 4.50 |
| GLIDE | | 0.28 | 3.21 | 2.03 | Somewhat Agree | Neutral | 1.18 | 58.12 |
| DALLE3 | | 0.29 | 3.23 | 3.65 | Somewhat Agree | Somewhat Agree | 0.41 | 11.50 |
| Stable Diffusion | BERT | 0.0009 | 2.50 | 3.22 | Neutral | Somewhat Agree | 0.72 | 22.36 |
| DALLE2 | | -0.0090 | 2.47 | 3.11 | Neutral | Somewhat Agree | 0.63 | 20.57 |
| GLIDE | | -0.0221 | 2.44 | 2.03 | Neutral | Neutral | 0.41 | 20.19 |
| DALLE3 | | -0.0066 | 2.48 | 3.65 | Neutral | Somewhat Agree | 1.17 | 47.17 |

*5.3.3 Analysis of Text-Image Alignment Metrics.* The CLIP Score as shown in Table 22 varying degrees of alignment with human judgments. For Stable Diffusion, the MAPE was exceptionally low at 1.86%, indicating a near-perfect correlation with human scores. DALLE2 also performed well with a MAPE of 4.50%, suggesting a close match. However, GLIDE had a high MAPE of 58.12%, reflecting significant misalignment. DALLE3's MAPE was 11.50%, indicating a moderate alignment. These results suggest that while the CLIP Score generally aligns well with human judgments for certain models, its performance can vary significantly. The BERT Score also exhibited varied accuracy compared to human judgments. For Stable Diffusion, the MAPE was 22.36%, indicating a moderate misalignment. DALLE2 showed a MAPE of 20.57%, reflecting a similar level of discrepancy. GLIDE had a MAPE of 20.19%, showing a relatively better alignment compared to CLIP Score but still significant divergence. DALLE3 had the highest MAPE of 47.17%, indicating a substantial misalignment. These results indicate that the BERT Score may not be as reliable in capturing human judgments of text-image alignment. Overall, it is very clear the CLIP score is close to human is all three cases expect GIIDE model, which is showing and its reliability.

## 6 CONCLUSION

In this paper, we presented an extensive analysis of generative images, encompassing various dimensions including photorealism, image quality, and text-image alignment. We developed a comprehensive questionnaire to evaluate these three key dimensions. The results of the human survey indicated that camera-generated images consistently outperformed AI-generated counterparts in photorealism and text-image alignment, highlighting the inherent challenges faced by current AI models in these areas. Among the AI models, DALL-E3 demonstrated significant advancements in image quality, surpassing even camera-generated images in certain aspects. However, it struggled with photorealism, indicating a trade-off between quality and realism.

The introduction of the Interpolative Binning Scale (IBS) represents a significant step forward in bridging the gap between metric scores and human evaluations. This method improves interpretability and provides a more nuanced understanding of how well computational metrics reflect human perceptions. Our analysis revealed that metrics like LPIPS and MS-SSIM showed closer alignment with human judgments, whereas FID and KID often deviated significantly, particularly for models like GLIDE. We also proposed the Neural Feature Similarity Score (NFSS), which outperformed all existing metrics in terms of assessing the quality of generative images.

These findings have profound implications for the development and deployment of AI-generated content. As the market for AI-generated images continues to expand, the need for robust, reliable, and human-aligned evaluation metrics becomes increasingly critical. This study provides a foundation for future research to refine and develop metrics that better capture the nuances of human visual perception, thereby enhancing the overall quality and reliability of AI-generated images. In conclusion, while significant progress has been made, the journey toward achieving perfect alignment between AI-generated images and human perceptual standards is ongoing. Our study highlights both the advancements and the challenges, providing a clear direction for future research and development in this rapidly evolving field.